\documentclass[aps,prx,floatfix,twocolumn,showpacs,nofitinbib,superscriptaddress,hyperref=pdftex,citeautoscript]{revtex4-2}

\usepackage{lmodern}

\usepackage{epsfig}
\usepackage{amsmath}
\usepackage{amssymb}
\usepackage{amsfonts}
\usepackage{ dsfont }
\usepackage{ comment,color }
\usepackage{lipsum}
\usepackage{hyperref,mathtools}
\usepackage{makecell}
\usepackage{array}
\usepackage{physics}
\usepackage{xcolor}
\usepackage[sep=3pt, offset=1em]{simpler-wick}
\usepackage{tikz}
\usetikzlibrary{quantikz}

\usepackage{siunitx}

\hypersetup{colorlinks=true,linkcolor=blue,anchorcolor=blue,citecolor=blue,filecolor=blue,urlcolor=blue,bookmarksnumbered=true,pdfview=FitB}

\newcommand{\beqa}{\begin{eqnarray}}
\newcommand{\eeqa}{\end{eqnarray}}

\usepackage[normalem]{ulem}
\renewcommand\sout{\bgroup\markoverwith{\textcolor{red}{\rule[0.5ex]{2pt}{0.4pt}}}\ULon}

\begin{document}

\hsize\textwidth\columnwidth\hsize\csname@twocolumnfalse\endcsname

\title{Entangling transmons with low-frequency protected superconducting qubits}

\author{Andrea Maiani, Morten Kjaergaard, and Constantin Schrade}
\affiliation{Center for Quantum Devices, Niels Bohr Institute, University of Copenhagen, 2100 Copenhagen, Denmark}

\date{\today}

\vskip1.5truecm
\begin{abstract}
Novel qubits with intrinsic noise protection constitute a promising route for improving the coherence of quantum
information in superconducting circuits. However, many protected superconducting qubits exhibit relatively low transition frequencies, which could make their integration with conventional transmon circuits challenging. In this work, we propose and study a scheme for entangling a
tunable transmon with a Cooper-pair parity-protected qubit, a paradigmatic example of a low-frequency protected qubit that stores quantum information in opposite Cooper-pair parity states on a superconducting island. By tuning the external flux on the transmon, we show that non-computational states can mediate a two-qubit entangling gate that preserves the Cooper-pair parity independent of the detailed pulse sequence. Interestingly, the entangling gate bears similarities to a controlled-phase gate in conventional transmon devices. Hence, our results suggest that standard high-precision gate calibration protocols could be repurposed for operating hybrid qubit devices.
\end{abstract}

\maketitle
Superconducting transmon qubits~\cite{koch2007} are a highly promising platform for noisy intermediate-scale quantum (NISQ) devices~\cite{preskill2018} and error-corrected quantum computers~\cite{Andersen2020,Marques2021,Krinner2021,Chen2021} with applications ranging from quantum simulations~\cite{arute2020,mi2021a,braumuller2021,karamlou2021} to the first experiments on quantum advantage~\cite{arute2019,wu2021}.  Among the most attractive features of the transmon circuit are its reproducibility, insensitivity to charge noise-induced dephasing, and coherence times that have seen steady improvements over the past decade~\cite{kjaergaard2020}.  Interestingly though, despite notable advances in prolonging the coherence of transmons, the transmon circuit does not exhibit \textit{intrinsic protection} to qubit relaxation errors. It is thus an important question how transmon devices can be further optimized with complementary qubit technologies to accelerate the path to fault-tolerant quantum computation.

Motivated by the challenge of exploring complementary qubit modalities, several alternative qubit encodings have been proposed~\cite{kitaev2006,weiss2019,brooks2013,dempster2014,Groszkowski2018,Paolo2019,Hyart2013,Landau2016,Plugge2016,Plugge2017,karzig2017,hoffman2016,schrade2018a,schrade2018b,blatter2001, ioffe2002b,doucot2002,doucot2005,gladchenko2008,doucot2012,gyenis2021b,smith2020,schrade2021} and experimentally studied~\cite{larsen2020,smith2020b,bell2014,gyenis2021a,kalashnikov2020,campagneibarcq2020,Manucharyan2009,Earnest2018,Lin2018,Nguyen2019,Hazard2019,Somoroff2021,Hassani2022}.
A particular class of such novel qubit encodings are Cooper-pair parity-protected qubits (PPQ) \cite{gyenis2021b,smith2020,schrade2021,larsen2020,smith2020b}, which rely on a special Josephson element that only permits tunneling of \textit{pairs} of Cooper-pairs. Similar to the transmon qubit, the two nearly-degenerate ground states of the PPQ have a nearly flat charge dispersion, which makes them insensitive to charge-noise induced dephasing. Similar to the fluxonium qubit \cite{Manucharyan2009,Earnest2018,Lin2018,Nguyen2019,Hazard2019,Somoroff2021}, the two qubit states also have disjoint support, since they carry opposite Cooper-pair parity. This disjoint support ensures that, if the qubit-environment coupling conserves the Cooper-pair parity, relaxation errors between the computational states are prevented.

While considerable efforts have been devoted to the development of a gate set for protected superconducting qubits~\cite{brooks2013,klots2021,Abdelhafez2020,Nesterov2018,Ficheux2021,Nesterov2021} for their use as an independent quantum computing platform, a different approach is to integrate protected qubits as memory elements in a conventional transmon-based quantum computing architecture.
Such \textit{heterogeneous} superconducting quantum processing architectures have recently gained increased attention~\cite{ciani2022}.
In such a scheme, the qubit state would be stored on the protected qubit during idle times and transferred to the transmon qubits for fast, high-fidelity operations, using the full machinery of well-established high-fidelity transmon operation. However, many protected qubits exhibit relatively low qubit transition frequencies, which could make this integration with transmon devices challenging. This motivates the question of efficiently generating entanglement between protected superconducting qubits and transmon qubits. 

In this work, we propose and study a capacitive coupling scheme for entangling a tunable transmon with a PPQ, a paradigmatic example of a protected superconducting qubit featuring a nearly-flat charge dispersion and qubit states with disjoint support. 
By tuning the external flux on the transmon, we show that non-computational states can mediate an entangling gate that 
preserves the Cooper-pair parity irrespective of the detailed pulse sequence. Besides opening the way to coherent state transfer, the proposed entangling gate also bears similarities with a controlled-phase gate in conventional capacitively coupled transmon qubits. Consequently, our results suggest standard high-precision two-qubit calibration protocols could be repurposed for the operation of hybrid qubit devices.

\section{Setup}
\begin{figure}[!t] \centering
\includegraphics[width=\linewidth]{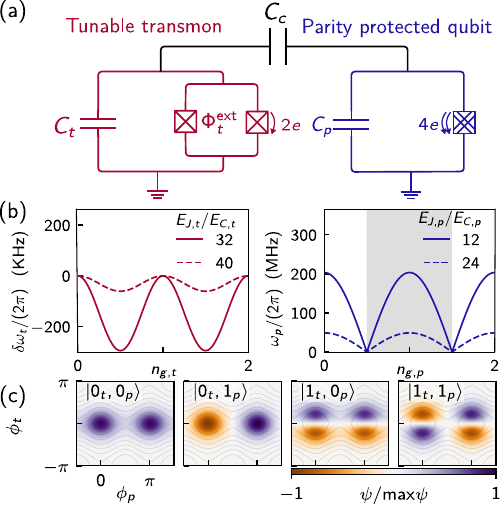}
\caption{\textbf{Hybrid qubit setup.}
(a) A frequency-tunable transmon (red) coupled to a PPQ (blue) realized by a capacitively-shunted tunneling element for pairs of Cooper-pairs (Josephson junction symbol with double lines). The two qubits are coupled via a coupling capacitance $C_c$. 
(b) Zoom-in on the charge dispersion relation for the transmon (left panel) and the PPQ (right panel). In the gray area, the ground (first excited) state carries odd (even) Cooper-pair parity. In the white area, the order is inverted.
(c) Wavefunctions of the decoupled system $(C_c=0)$. The system parameters are $(E_{J,t}, E_{J,p},E_{C,t},E_{C,p}) 
= 2\pi (10,3,0.25,0.25)\,\text{GHz}$.
}\label{fig:1}
\end{figure}

As depicted in Fig.\,\ref{fig:1}(a), we consider a direct capacitively coupling between a frequency-tunable transmon qubit and 
a PPQ, realized by a capacitively-shunted $\cos(2\phi)$ Josephson element for the tunneling of pairs of Cooper-pairs. The individual Hamiltonians of the transmon, $H_{t}$, 
and of the PPQ, $H_{p}$, are given by, 
\begin{equation}
\begin{split}
    H_{t} &= 4 E^{}_{C,t} (n_{t} - n^{}_{g,t})^2 - E^{}_{J,t}\, \text{cos}(\phi_{t}) \,, \\
    H_{p} &= 4 E^{}_{C,p} (n_{p} - n_{g,p})^2 - E^{}_{J,p} \, \text{cos}(2\phi_{p})\,.
\end{split}
\label{Eq1}
\end{equation}
Here, $(n_{t},\phi_{t})$ and $(n_{p},\phi_{p})$ denote Cooper-pair charge and phase degrees of freedom of the transmon and PPQ. Moreover, $E_{J,t}$ is the transmon Josephson energy and $E_{J,p}$ is the two-Cooper-pair tunneling amplitude of the PPQ. The charging energies of the two qubit circuits are $E^{}_{C,t} =e^{2}/2C_{t}$ and $E^{}_{C,p} =e^{2}/2C_{p}$ with the shunt capacitances $C_{t}$ and $C_{p}$.

Both Hamiltonians in Eq.\,\eqref{Eq1} can be diagonalized exactly by rewriting the eigenvalue problems as Mathieu equations. For the transmon \cite{koch2007}, the energy splitting between the ground and first-excited state, which form the qubit basis $\left|0_{t}\right\rangle$ and $\left|1_{t}\right\rangle$, is $\omega_{t}=\sqrt{8E_{J,t}E_{C,t}}+\delta\omega_{t}$ with $\delta \omega_t\propto \text{exp}(-\sqrt{8E_{J,t}/E_{C,t}}) \cos(2\pi n_{g,t})$  for $E_{J,t}\gg E_{C,t}$, see the left panel of Fig.\,\ref{fig:1}(b). Here and in the following discussions, we have put $\hbar=1$. For the PPQ \cite{smith2020}, the qubit basis is given by the two lowest-energy states with even and odd Cooper-pair parity, $\left|0_{p}\right\rangle$ and $\left|1_{p}\right\rangle$. These states have an exponentially suppressed energy splitting, $\omega_{p}\propto \text{exp}(-\sqrt{2E_{J,p}/E_{C,p}}) |\cos(\pi n_{g,p})|$ for $E_{J,p}\gg E_{C,p}$, see the right panel of Fig.\,\ref{fig:1}(b). Unlike the transmon, the PPQ is thus a low-frequency qubit with the computational states exhibiting an exact degeneracy if $\text{cos}(\pi n_{g,p})=0$ and a near-degeneracy otherwise. However, like the transmon, the energy splitting of the PPQ is insensitive to variations in $n_{g,p}$ if $E_{J,p}\gg E_{C,p}$, which ensures insensitivity to charge noise dephasing.

To explain the protection of the PPQ against parity-preserving relaxation errors, we consider the wavefunctions of the computational basis. In phase space, these wavefunctions are symmetric/anti-symmetric combinations of states that are localized in the $0$- and $\pi$-valleys of the Josephson potential, see Fig.\,\ref{fig:1}(c). In charge space, the same wavefunctions are superpositions of states with even/odd Cooper-pair number. Due to this disjoint support of the charge space wavefunctions, $\left\langle 0_{p} | \mathcal{O} | 1_{p}\right\rangle = 0$ for any operator $\mathcal{O}$ that preserves the Cooper-pair parity, which is the condition for protection against parity-preserving relaxation errors \cite{gyenis2021b}. 

\begin{figure*}[!htp]
\centering
\includegraphics[width=\linewidth]{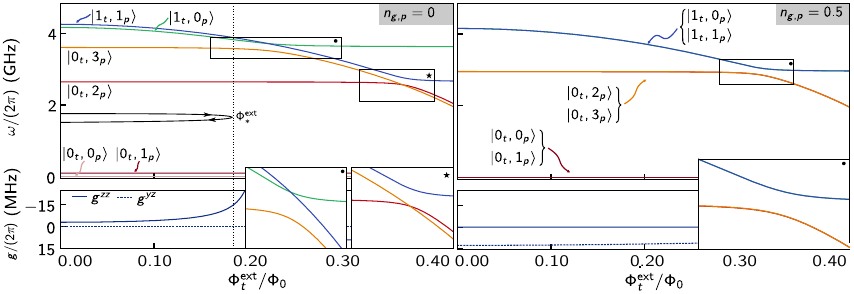}
\caption{
\textbf{Low-energy spectrum of the hybrid qubit setup and $\textsf{CZ}_\phi$ gate.}
(a) Low-energy spectrum of the hybrid qubit setup for $n_{g, p}=0$ and $(E_{J,t}, E_{J,p},E_{C,t},E_{C,p},E_{C,c}) 
= 2\pi (12,2.7,0.2,0.18,0.025)\,\text{GHz}$ as a function of the external flux $\Phi_{\text{ext}}$ of the tunable transmon. 
The $\textsf{CZ}_\phi$ gate is realized by a rapid excursion from $\Phi_{\text{ext}}=0$ to the vicinity of the 
$\left|1_{t},0_{p}\right\rangle\leftrightarrow\left|0_{t},3_{p}\right\rangle$ anti-crossing at  $\Phi_{\text{ext}}=\Phi^{*}_{\text{ext}}$. The bottom panel shows the coupling strengths, $g^{yz}$ and $g^{zz}$, upon approaching the anti-crossing. 
(b) Same as (a) but for $n_{g, p}=0.5$. Each shown energy level is now exactly two-fold degenerate. 
}\label{fig:2}
\end{figure*}

Having introduced the two decoupled qubit circuits with the associated computational subspace $\mathcal{P}_0=
\{
\left|1_{t},1_{p}\right\rangle,
\left|1_{t},0_{p}\right\rangle,
\left|0_{t},1_{p}\right\rangle,
\left|0_{t},0_{p}\right\rangle
\}
$, we proceed by coupling the qubits via a standard capacitive coupling (see also Fig.~\ref{fig:1}a) corresponding to a coupling Hamiltonian given by
\begin{equation}
    H_{c} = 4 E_{C,c} (n_{p} - n_{g,p})(n_{t} - n_{g,t}).
\label{Eq2}
\end{equation}
Here, $E_{C,c} = e^2 C_{c}/(C_{p}C_{t})$ with the coupling capacitance $C_{c}$. We note that there are two main advantages of the proposed capacitive coupling: (1) It always conserves the Cooper-pair parity on the PPQ, ensuring the PPQ's protection against relaxation errors. (2) It is compatible with standard transmon technology and is routinely used for coupling transmons in state-of-the-art architectures \cite{kjaergaard2020}. Despite these possible advantages of a capacitive coupling, we acknowledge, though, that inductively couplings have also been studied for entangling transmons, such as the `gmon' circuit \cite{Chen2014}. However, since the inductive coupling in the `gmon' arises from a conventional Josephson junction, we expect that maintaining the Cooper-pair parity conservation could be a challenge when generalizing such a scheme to a PPQ/transmon hybrid system.

To conclude, the full Hamiltonian of our setup is $H_{}=H_{p}+H_{t}+H_{c}$. 
In the next section, we will derive the effective qubit interaction due to this direct capacitive coupling.

\section{Effective Hamiltonian}

To motivate the derivation of the effective qubit interaction, we first recall the case of two capacitively coupled transmon qubits, $t1$ and $t2$, which are both `high-frequency' qubits. In this example, the capacitive coupling mediates a $\sigma_{t1}^{y}\sigma_{t2}^{y}$ interaction when projected onto the computational subspace and a $\sigma_{t1}^{z}\sigma_{t2}^{z}$ interaction due to the mixing of computational and non-computational states~\cite{krantz2019}. In our setup, which involves the coupling of a `high-frequency' transmon qubit and a `low-frequency' PPQ, we will show that the couplings to non-computational states will play an even more essential role. To anticipate this result, we note that the coupling Hamiltonian of Eq.\,\eqref{Eq2} at $n_{g,p}=0$ vanishes exactly when projected onto the computational subspace,  $\langle s_{t}, s'_{p}|H_{c}|s''_{t},s'''_{p}\rangle=0$ for any two states $|s_{t},s'_{p}\rangle, |s''_{t},s'''_{p}\rangle$ in $\mathcal{P}_0$ since $\langle 0_{t}|n_{t}|0_{t}\rangle=\langle 1_{t}|n_{t}|1_{t}\rangle=0$ and $\langle 0_{p}|n_{p}|1_{p}\rangle=0$. A direct coupling within the computational subspace is thus fully absent at $n_{g,p}=0$ and any qubit interaction, if present, is necessarily mediated by virtual transitions through non-computational states. 

\subsection{Special case: $\boldsymbol{n_{g,p}}=0$}

To identify the origin of such virtual transitions, we initially compare two special cases with the offset charge on the PPQ set to either $n_{g,p}=0$ or $n_{g,p}=0.5$. Starting with the $n_{g,p}=0$ case, we show the low-energy spectrum as a function of external magnetic flux $\Phi^{\text{ext}}_{t}$ of the tunable transmon in Fig.\,\ref{fig:2}(a). The spectrum comprises not only the four qubit levels of $\mathcal{P}_0$ but also two additional levels corresponding to the $\left|0_{t},2_{p}\right\rangle$ and $\left|0_{t},3_{p}\right\rangle$ state of the uncoupled system. 
Interestingly, the \textit{non-computational} states exhibit two anti-crossings with the \textit{computational} states, $\left|1_{t},1_{p}\right\rangle\leftrightarrow\left|0_{t},2_{p}\right\rangle$ and $\left|1_{t},0_{p}\right\rangle\leftrightarrow\left|0_{t},3_{p}\right\rangle$, at certain values of external flux. These anti-crossings arise because the respective couplings preserve the Cooper-pair parity on the PPQ. On the other hand, anti-crossing between $\left|1_{t},0_{p}\right\rangle\leftrightarrow\left|0_{t},2_{p}\right\rangle$ and $\left|1_{t},1_{p}\right\rangle\leftrightarrow\left|0_{t},3_{p}\right\rangle$ are absent from the spectum in Fig.\,\ref{fig:2}(a), as such couplings violate the Cooper-pair parity conservation on the PPQ. We will now show that in the vicinity of the two anti-crossings, virtual transitions in-and-out of the computational subspace are enhanced and, consequently, can induce a sizable effective qubit interaction between the transmon and the PPQ.

For computing the effective qubit interaction at $n_{g,p}=0$,  we initially project our setup Hamiltonian $H$ onto the four qubit states of $\mathcal{P}_0$ and on the additional $\left|0_{t},2_{p}\right\rangle$ and $\left|0_{t},3_{p}\right\rangle$ states. This yields the following low-energy Hamiltonian, 
\begin{equation}
H^{(n_{g,p}=0)}_{\text{low}}
=
\begin{pmatrix}
\omega_{11} & 0 & 0 & 0 &\lambda' & 0 \\
0 & \omega_{10}  & 0 & 0 & 0 & -\lambda'' \\
0 & 0 & \omega_{01} & 0 & 0 & 0 \\
0 & 0 & 0 & \omega_{00} &  0& 0  \\
\lambda'  & 0 & 0 & 0 & \omega_{02} & 0 \\
0 & -\lambda''  & 0 & 0 & 0 & \omega_{03}\\
\end{pmatrix}
.
\label{lowenergy1}
\end{equation}
Here, the frequency of the the $\left|s_{t},s'_{p}\right\rangle$ state in the uncoupled system is denoted by $\omega_{ss'}=\omega_{t,s}(\Phi^\mathrm{ext}_t)+\omega_{p,s'}$. Moreover, the coupling matrix elements are given by $\lambda'=\left\langle 1_{t},1_{p}|H_{c}|0_{t},2_{p}\right\rangle$ and $\lambda''=\left\langle 1_{t},0_{p}|H_{c}|0_{t},3_{p}\right\rangle$, where we picked a wavefunction gauge for which ($\lambda',\lambda''$) are real-valued. We point out that the low-energy Hamiltonian of Eq.\,\eqref{lowenergy1} is different from the one of capacitively-coupled transmons \cite{krantz2019}, because the conservation of Cooper-pair parity prohibits a coupling of the $\left|0_{t},1_{p}\right\rangle$ to the $\left|1_{t},0_{p}\right\rangle$ state. Also, for two coupled transmons only the highest energy computational state exhibits crossing with non-computational states. In our case, the two computational states, $\left|1_{t},0_{p}\right\rangle$ and $\left|1_{t},1_{p}\right\rangle$, both cross with non-computational states, albeit at different values of external flux.

Next, we integrate out the non-computational states to second order in $\lambda'$ and $\lambda''$ by a Schrieffer-Wolff transformation \cite{Bravyi2011_SM}. Provided that $\lambda'^{2}\ll |\omega_{02}-\omega_{11}|$ and $\lambda''^{2}\ll |\omega_{03}-\omega_{10}|$, we find that the effective qubit Hamiltonian reads \cite{bib:supplemental}, 
\begin{align}
H^{(n_{g,p}=0)}_{\text{eff}}
&=
\left(\omega_{t}+\frac{g^{zz}_{+}}{2}\right)
\frac{
\sigma^{z}_{t}
}
{
2
}
+
\left(\omega_{p}+\frac{g^{zz}_{-}}{2}\right)
\frac{
\sigma^{z}_{p}
}
{
2
}
+
g^{zz}_{-}\,\frac{
\sigma^{z}_{t}
}
{
2
}
\frac{
\sigma^{z}_{p}
}
{
2
},\nonumber
\\
g^{zz}_{\pm}&=\frac{\lambda'^{2}}{\omega_{11}-\omega_{02}}\pm\frac{\lambda''^{2}}{\omega_{10}-\omega_{03}},
\label{Heffng0}
\end{align}
where $\omega_{p/t}=\omega_{p/t,1}-\omega_{p/t,0}$ denote the bare qubit frequencies. The key insight from Eq.\,\eqref{Heffng0} is that the interaction between the two qubits is of $\sigma^{z}_{p}\sigma^{z}_{t}$ type. As anticipated, this interaction arises from a two-step perturbative sequence involving virtual transitions in-and-out of the 
$\left|0_{t},2_{p}\right\rangle$ and $\left|0_{t},3_{p}\right\rangle$ state. For example, in a perturbative sequence close to the $\left|1_{t},1_{p}\right\rangle\leftrightarrow\left|0_{t},2_{p}\right\rangle$ anti-crossing, the system exhibits a first virtual transition from the computational state $\left|1_{t},1_{p}\right\rangle$ to the non-computational state $\left|0_{t},2_{p}\right\rangle$ and, subsequently, a second virtual transition back to $\left|1_{t},1_{p}\right\rangle$. Such a sequence preserves the state of the transmon, which explains why the interaction is $\propto\sigma^{z}_{t}$. The dependence of the interaction on $\sigma^{z}_{p}$ arises because the coupling Hamiltonian of Eq.\,\eqref{Eq2} preserves the Cooper-pair parity.

\subsection{Special case: $\boldsymbol{n_{g,p}}=0.5$}

Having derived the qubit interaction at $n_{g,p}=0$, we want to compare the results of Eq.\,\eqref{Heffng0} with the $n_{g,p}=0.5$ case. We therefore plot the low-energy spectrum at $n_{g,p}=0.5$ in Fig.\,\ref{fig:2}(b). Unlike in the previous case, we find that each depicted energy level exhibits an exact two-fold degeneracy, corresponding to opposite Cooper-pair parity sectors. This finding is consistent with our results of Fig.\,\ref{fig:1}(b), where we pointed out
that the levels on the uncoupled PPQ are exactly degenerate at $n_{g,p}=0.5$. In particular, the anti-crossings $\left|1_{t},1_{p}\right\rangle\leftrightarrow\left|0_{t},2_{p}\right\rangle$ and $\left|1_{t},0_{p}\right\rangle\leftrightarrow\left|0_{t},3_{p}\right\rangle$ occur now at the same value of external flux and overlap exactly. Couplings between $\left|1_{t},0_{p}\right\rangle\leftrightarrow\left|0_{t},2_{p}\right\rangle$ and $\left|1_{t},1_{p}\right\rangle\leftrightarrow\left|0_{t},3_{p}\right\rangle$ remain absent (they are forbidden since the states belong to a different parity sector). We will now show that this new scenario at $n_{g,p}=0.5$ will lead to a different effective qubit Hamiltonian compared to Eq.\,\eqref{Heffng0}.

We begin again by projecting the setup Hamiltonian $H$ onto the qubit subspace $\mathcal{P}_{0}$ and onto the states $\left|0_{t},2_{p}\right\rangle$ and $\left|0_{t},3_{p}\right\rangle$. The resulting low-energy Hamiltonian reads,
\begin{equation}
H^{(n_{g,p}=0.5)}_{\text{low}}
=
\begin{pmatrix}
\omega_{11} & 0 & -i\eta & 0 & \lambda & 0 \\
0 & \omega_{10}  & 0 & i\eta & 0 & -\lambda \\
 i\eta  & 0 & \omega_{01} & 0 & 0 & 0 \\
0 & -i\eta & 0 & \omega_{00} & 0 & 0  \\
\lambda & 0 & 0 & 0 & \omega_{02} & 0 \\
0& -\lambda & 0 & 0& 0 & \omega_{03}
\end{pmatrix}.
\end{equation}
Here, we have $\lambda=\left\langle 1_{t},1_{p}|H_{c}|0_{t},2_{p}\right\rangle=-\left\langle 1_{t},0_{p}|H_{c}|0_{t},3_{p}\right\rangle$ and $\eta=i\left\langle 1_{t},1_{p}|H_{c}|0_{t},1_{p}\right\rangle=-i\left\langle 1_{t},0_{p}|H_{c}|0_{t},0_{p}\right\rangle$ in a wavefunction gauge for which $(\lambda,\eta)$ are real-valued. By inserting the coupling Hamiltonian in the expressions for the matrix elements, we note that $\eta\propto\left\langle s_{p}|n_{p}|s_{p}\right\rangle-n_{g,p}$. In the previous case when $n_{g,p}=0$, we had $\left\langle s_{p}|n_{p}|s_{p}\right\rangle=0$ and, consequently, $\eta$ vanished. In the present case when $n_{g,p}=0.5$ and $E_{J,p}\gtrsim E_{C,p}$, we have $\left\langle s_{p}|n_{p}|s_{p}\right\rangle\neq n_{g,p}$ so that $\eta$ is finite yet gets successively smaller upon increasing $E_{J,p}$. 
In particular, when $E_{J,p}\gg E_{C,p}$, we have $\left\langle s_{p}|n_{p}|s_{p}\right\rangle\rightarrow n_{g,p}$ so that the contribution of $\eta$ to the low-energy Hamiltonian is negligible. 
Lastly, we note that due to the degeneracy of the PPQ levels, $\omega_{p,0}=\omega_{p,1}$ and $\omega_{p,2}=\omega_{p,3}$. Hence, the frequencies of the hybrid setup satisfy $\omega_{11}=\omega_{10}$, $\omega_{01}=\omega_{00}$, and $\omega_{02}=\omega_{03}$.

We now proceed by integrating out the effects of the non-computational states, $\left|0_{t},2_{p}\right\rangle$ and $\left|0_{t},3_{p}\right\rangle$, with a Schrieffer-Wolff transformation. The resulting effective Hamiltonian is of the form,
\begin{equation}
\begin{split}
H^{(n_{g,p}=0.5)}_{\text{eff}}&=
\left(\omega_{t}+\frac{g^{zz}_{+}}{2}\right)
\frac{
\sigma^{z}_{t}
}
{
2
}
+
g^{yz}\,\sigma^{y}_{t}\sigma^{z}_{p},\\
g^{yz}&=\eta,
\end{split}
\end{equation}
Contrasting this result with Eq.\,\eqref{Heffng0}, we note that both terms $\propto\sigma^{z}_{p}$ and $\propto\sigma^{z}_{t}\sigma^{z}_{p}$ have vanished because $\omega_{p}=0$ and $g^{zz}_{-}=0$. As a result, the effective qubit interaction is \textit{not} of $\sigma^{z}_{t}\sigma^{z}_{p}$ but rather of $\sigma^{y}_{t}\sigma^{z}_{p}$ type. The physical origin of the interaction at $n_{g,p} = 0.5$ is different from the $n_{g,p} = 0$ case, since it arises directly from the finite matrix elements of the charge operators of the parity-protected qubit and the transmon qubits. The non-computational states induce only a renormalization of the transmon frequency through the $g^{zz}_{+}$ contribution in the coefficient of $\sigma^{z}_{t}$.

\begin{figure}[!t] \centering
\includegraphics[width=1.\linewidth] {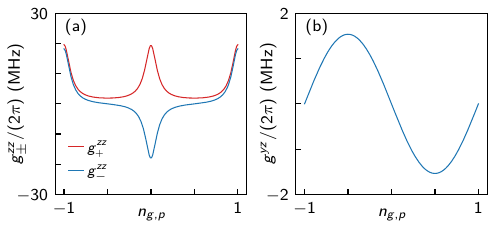}
\caption{\textbf{Gate charge dependence of the two-qubit couplings.}
(a) Couplings, $g^{zz}_{+}$ (red) and $g^{zz}_{-}$ (blue), versus the gate charge, $n_{g,p}$, on the PPQ. 
The system parameters are $(E_{J,t},E_{J,p},E_{C,t},E_{C,p},E_{C,c}) = 2\pi (12,2.7,0.2,0.15,0.005)\,\text{GHz}$ and $\Phi/\Phi_{0}=0.282$. 
The curves are $2$-periodic in $n_{g,p}$. This periodicity results because for $|n_{g,p}|<0.5$, the even and odd Cooper-pair parity states,
$|0_{p}\rangle$ and $|1_{p}\rangle$, are the ground and first-excited state, while for $0.5<|n_{g,p}|<1$ this ordering is reversed so that the odd Cooper-pair parity state $|1_{p}\rangle$ is the ground state and the even Cooper-pair parity state $|0_{p}\rangle$ is the first-excited state.
(b) Coupling, $g^{yz}$ (blue), versus the gate charge, $n_{g,p}$, on the PPQ. The system parameters are the same as in (a). 
For the chosen parameters, $g^{y}\ll 1\,\text{MHz}$ and, therefore, this single-qubit term is not shown.
\label{figure6}}
\end{figure}

\subsection{General case}

So far, we have seen that the capacitive coupling between the transmon and the PPQ induces a qubit interaction that is substantially different for $n_{g,p}=0$ and $n_{g,p}=0.5$. In a last step, we want to interpolate between those two representative cases. This interpolation is achieved by studying the dependence on the offset charge $n_{g,p}$ of the various matrix elements. Since the procedure for obtaining the effective interaction is otherwise identical to the special cases, we only note that for generic values of $n_{g,p}$ the effective Hamiltonian acquires both a $\sigma^{z}_{t}\sigma^{z}_{p}$ and a $\sigma^{y}_{t}\sigma^{z}_{p}$ interaction term \cite{bib:supplemental}, 
\begin{align}
H^{(n_{g,p})}_{\text{eff}}&=
\left(\omega_{t}+\frac{g^{zz}_{+}}{2}\right)
\frac{
\sigma^{z}_{t}
}
{
2
}
+
\left(\omega_{p}+\frac{g^{zz}_{-}}{2}\right)
\frac{
\sigma^{z}_{p}
}
{
2
}
+
g^{y}_{t}
\sigma^{z}_{t}
\nonumber
\\
&+
g^{zz}_{-}\,\frac{
\sigma^{z}_{t}
}
{
2
}
\frac{
\sigma^{z}_{p}
}
{
2
}
+
g^{yz}\,\sigma^{y}_{t}\sigma^{z}_{p}.
\end{align}
Here, the couplings are given by,
\begin{equation}
\begin{split}
g^{zz}_{\pm}&=
\frac{\lambda'^{2}}{\omega_{11}-\omega_{02}}
\pm
\frac{\lambda''^{2}}{\omega_{10}-\omega_{03}}
\\
g^{yz}&=(\eta'+\eta'')/2,\\
g^{y}_{t}&=(\eta'-\eta'')/2,
\end{split}
\label{gs}
\end{equation}
where $\lambda'=\left\langle 1_{t},1_{p}|H_{c}|0_{t},2_{p}\right\rangle$, $\lambda''=\left\langle 1_{t},0_{p}|H_{c}|0_{t},3_{p}\right\rangle$, $\eta'=i\left\langle 1_{t},1_{p}|H_{c}|0_{t},1_{p}\right\rangle$ and $\eta''=-i\left\langle 1_{t},0_{p}|H_{c}|0_{t},0_{p}\right\rangle$. With the help of Eq.\,\eqref{gs}, we are now in the position to numerically evaluate the couplings as a function of $n_{g,p}$. As shown in Fig.\,\ref{figure6}, the transition from a a pure $\sigma^{z}_{t}\sigma^{z}_{p}$ at $n_{g,p}=0$ to a pure $\sigma^{y}_{t}\sigma^{z}_{p}$ at $n_{g,p}=0.5$ is gradual. Moreover, while the the functional form of $g^{zz}_{\pm}(n_{g,p})$ is more complicated, we find that the functional form of $g^{yz}(n_{g,p})$ is approximately sinusoidal, $g^{yz}(n_{g})\approx g^{yz}_{0}\sin(\pi n_{g,p})$.
 As for the the dependence on the transmon offset charge, we remark that in the deep-transmon regime,  $E_{J,t}\gg E_{C,t}$,  the qubit interaction is almost independent of $n_{g,t}$.

So far, we have derived the effective qubit interaction and have demonstrated that it depends on the anti-crossings with the non-computational states, $\left|0_{t},2_{p}\right\rangle$ and $\left|0_{t},3_{p}\right\rangle$. To realize the respective anti-crossings, 
we note that it is essential that,
\begin{equation}
\omega_{02}<\omega_{10}.
\label{condition}
\end{equation}
The transmon energy levels are approximated by $\omega_{t,n}\approx\sqrt{8E_{J,t}E_{C,t}}(n+1/2)-E_{J,t}$ while the PPQ energy levels by 
$\frac{\omega_{p,2} + \omega_{p,3}}{2} \approx2\sqrt{8E_{J,p}E_{C,p}}-4 E_{J,p}$. Neglecting
the anharmonicity corrections on both qubits, we find that the necessary condition in Eq.\,\eqref{condition} simplifies to $2\sqrt{E_{J,p}E_{C,p}}<\sqrt{E_{J,t}E_{C,t}}$. This condition is satisfied for the parameters chosen in Fig.\,\ref{fig:2}.

\section{Quantum Gates}
We will now use the effective Hamiltonian for the hybrid PPQ/transmon setup to implement a controlled-phase gate ($\textsf{CZ}_\phi$), which will preserve the Cooper-pair parity irrespective of the detailed pulse sequence. In addition, we will also discuss a complete set of single-qubit gates realized by controllably driving the system in-and-out of protection \cite{kalashnikov2020}. In combination with the $\textsf{CZ}_\phi$ gate, these single-qubit gates will permit the coherent state transfer, a \textsf{SWAP} operation, between the transmon and PPQ.

\subsection{$\textsf{CZ}^{10}_\phi$ gate}

For deriving the $\textsf{CZ}^{10}_\phi$ gate protocol, we initially move to the frame that rotates with the bare qubit frequencies, $\tilde{H}^{(n_{g,p})}_{\text{eff}}=U^{\dag}(t)H^{(n_{g,p})}_{\text{eff}}U(t)-iU^{\dag}(t)\dot{U}(t)$ with 
$U(t)=e^{i(\omega_{t}\sigma^{z}_{t}
+
\omega_{p}\sigma^{z}_{p})t/2}$. Within this rotating frame, the effective Hamiltonian reads, 
\begin{equation}
\begin{split}
\tilde{H}^{(n_{g,p})}_{\text{eff}}&=
\frac{g^{zz}_{+}}{2}
\frac{
\sigma^{z}_{t}
}
{
2
}
+
\frac{g^{zz}_{-}}{2}
\frac{
\sigma^{z}_{p}
}
{
2
}
+
g^{zz}_{-}\,\frac{
\sigma^{z}_{t}
}
{
2
}
\frac{
\sigma^{z}_{p}
}
{
2
}
\\
&+[-ie^{i\omega_{t}t}
(
g^{yz}\,
\sigma^{+}_{t}\sigma^{z}_{p}
+
g^{y}\,
\sigma^{+}_{t})+\text{H.c.}],
\label{rotframeHamiltonian}
\end{split}
\end{equation}
where we introduced $\sigma^{\pm}_{}=(\sigma^{x}_{}\pm i\sigma^{y}_{})/2$. We note that the terms $\propto g^{yz}\sigma^{\pm}_{t}\sigma^{z}_{p}$ and $\propto g^{y}\sigma^{\pm}_{t}$ vanish if $n_{g,p}=0$. In this situation, the free evolution of the effective Hamiltonian can implement a $\textsf{CZ}^{10}_\phi$ gate. For executing this $\textsf{CZ}^{10}_\phi$ gate, we carry out a rapid excursion from $\Phi^{\text{ext}}_t\approx 0$ to a flux $\Phi^{\text{ext}}_t\approx \Phi^{\text{ext}}_{*}$ close to the anti-crossing $\left|1_{t},0_{p}\right\rangle\leftrightarrow\left|0_{t},3_{p}\right\rangle$. We then let the system evolve freely for a time $t_{*}=\phi$. This free evolution gives rise to a rotation in the space of $\left|1_{t},0_{p}\right\rangle$ and $\left|0_{t},3_{p}\right\rangle$. 
After the time $t_{*}$, the $\left|1_{t},0_{p}\right\rangle$ state will have acquired a finite phase factor and we rapidly return to the idle configuration at $\Phi^{\text{ext}}_t\approx 0$. Because $g^{zz}_{-}\approx-g^{zz}_{+}$ near the anti-crossing, the result of this rapid excursion is a $\textsf{CZ}^{10}_{\phi}$ gate of the form, 
\begin{equation}
\begin{split}
\textsf{CZ}^{10}_\phi
&=
|0_{t}\rangle\langle 0_{t}| \otimes \textsf{I}_{p}
+
|1_{t}\rangle\langle 1_{t}| \otimes \textsf{P}_{p}
\\
 \textsf{P}_{p}&=e^{-i\phi}|0_{p}\rangle\langle 0_{p}|+|1_{p}\rangle\langle 1_{p}|
 \end{split}
\end{equation}
Unlike for the case of two capacitively coupled transmons $t1$ and $t2$, we remark that the phase factor is not acquired by the $\left|1_{t1},1_{t2}\right\rangle$ state but by the $\left|1_{t},0_{p}\right\rangle$ state. Also, as announced at the beginning of this section, we highlight that the Cooper-pair parity is preserved for the full duration of the $\textsf{CZ}^{10}_\phi$ gate.

In the protocol for the $\textsf{CZ}^{10}_{\phi}$ gate, we have assumed that the offset charge on the PPQ is gate-tuned to $n_{g,p}=0$. Such a tuning is beneficial as it maximizes the coefficient of the $\sigma^{z}_{t}\sigma^{z}_{p}$ terms, thereby allowing for improved gate speed. Furthermore, the tuning should always be achievable because higher levels of the PPQ are strongly offset charge sensitive and can be used for adjusting $n_{g,p}$. However, the fine-tuning to $n_{g,p}=0$ is not essential for the gate protocol. To see this, we note that the terms $\propto g^{yz}\sigma^{\pm}_{t}\sigma^{z}_{p}$ and $\propto g^{y}\sigma^{\pm}_{t}$ in Eq.\,\eqref{rotframeHamiltonian}, which appear when $n_{g,p}$ is detuned from zero, share a fast-oscillating prefactor $\propto e^{i\omega_{t}t}$. This fast-oscillating prefactor suggests that such terms are average to zero when invoking a `rotating-wave approximation'. For making this argument more precise, we have integrated out the fast-oscillating terms to second order in $g^{y}$ and $g^{yz}$ with a time-dependent Schrieffer-Wolff transformation \cite{eckstein2017,petrescu2021}. The resulting modified effective Hamiltonian reads \cite{bib:supplemental},
\begin{align}
\tilde{H}^{(n_{g,p})}_{\text{eff}}(t)&\approx
\left(
\frac{g^{zz}_{+}}{2}
+
\frac{4[\tilde{g}^{y}(t)^{2}+\tilde{g}^{yz}(t)^{2}]}{\omega_{t}}
\right)
\frac{
\sigma^{z}_{t}
}
{
2
}
+
\frac{g^{zz}_{-}}{2}
\frac{
\sigma^{z}_{p}
}
{
2
}\nonumber
\\
&+
\left(
g^{zz}_{-}
+
\frac{16\tilde{g}^{y}(t)\tilde{g}^{yz}(t)}{\omega_{t}}
\right)
\,\frac{
\sigma^{z}_{t}
}
{
2
}
\frac{
\sigma^{z}_{p}
}
{
2
},
\label{rotframeHamiltonian2}
\end{align}
with $\tilde{g}(t)=g\sin(\omega_{t}t/2)$.
Provided that $g^{y}\ll \omega_{t}$ and $g^{yz}\ll \omega_{t}$, we see that the correction terms to the effective Hamiltonian are indeed negligibly small. For the realistic parameters chosen in Fig.\,\ref{fig:2}, we have $g^{y} / (2 \pi) =\SI{345}{\kilo\Hz}$ and $g^{yz}/ (2\pi) =\SI{3.88}{\mega\Hz}$ if $n_{g,p}=0.1$. 

\begin{figure}[!t] \centering
\includegraphics[width=1.\linewidth] {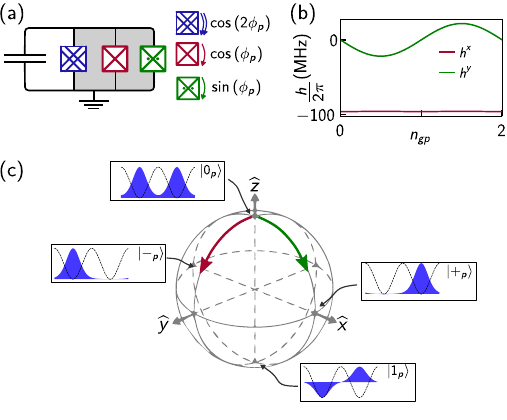}
\caption{\textbf{Single-qubit gates.}
(a) A generalized PPQ circuit with three circuits elements; a $\cos(2\phi_{p})$ element (blue), a $\cos(\phi_p)$ element (red), and a $\sin(\phi_p)$ element (green). No magnetic flux is threading the gray areas of the circuit. (b) Dependence of the matrix element $h^{x}$ (resulting from the $\cos(\phi_p)$ element) and $h^{y}$ (resulting from the $\sin(\phi_p)$ element) as a function of $n_{g,p}$. The system parameters are $(E_{J, p}, E_{C, p}) = 2\pi (2.7, 0.18)\,\text{GHz}$ and $(\varepsilon_x/E_{J, p},\varepsilon_y/E_{J, p})=(0.04,0.2)$. (c) The $\cos(\phi_p)$ element element induces rotations around the $x$-axis of the Bloch sphere (red arrow). The $\sin(\phi_p)$ element induces rotations around the $y$-axis of the Bloch sphere. The wavefunctions of the PPQ in the $z$-basis ($|0_{p}\rangle$,$|1_{p}\rangle$) and in the $x$-basis ($|+_{p}\rangle$,$|-_{p}\rangle$) are shown schematically. 
}\label{fig:3}
\end{figure}

\subsection{Single-qubit gates}
Having introduced the $\textsf{CZ}^{10}_\phi$ gate, we now discuss the implementation of single-qubit gates on the PPQ. For implementing these single-qubits gates, we consider the generalized circuit for a PPQ depicted in Fig.\,\ref{fig:3}(a). The circuit comprises not only a $\cos(2\phi_{p})$ element for the tunneling of pairs of Cooper-pairs, but also a $\cos(\phi_{p})$ and $\sin(\phi_{p})$ element that describe the tunneling of single Cooper-pairs.
The Hamiltonians for these additional circuit elements are given by,
\begin{equation}
\begin{split}
H^{x}_{}
&= 
-\varepsilon^x\,\cos(\phi_{p}),
\\
H^{y}_{}
&= 
-\varepsilon^{y}\,\sin(\phi_{p}), 
\label{single-qubit-perturbations}
\end{split}
\end{equation}
While both additional circuit elements permit single Cooper-pair tunnelings and temporarily lift the qubit protection, they are typically tuned by different control parameters, depending on the experimental implementation of the PPQ \cite{larsen2020,smith2020b}. For example, if the PPQ is realized in a nanowire Josephson interferometer, then sinusoidal term arises if the interferometer junctions are tuned out of balance by local gate electrodes. In contrast, the cosinusoidal term arise when the interferometer magnetic flux is biased away from half flux quantum \cite{schrade2021,larsen2020}.

We now project the Hamiltonians $H_{p}+H^{x}_{}$ and $H_{p}+H^{y}_{}$ onto the computational subspace of the PPQ. The resulting qubit Hamiltonians read,
\begin{align}
H^{x}_{\text{eff}}
&= 
\delta\omega_{p}\cos(\pi n_{g,p})\,\sigma^{z}_{p}\,/2
+
\delta h^{x}\,\sigma^{x}_{p},\nonumber
\\
H_{\text{eff}}^{y}
&= 
\delta\omega_{p}\cos(\pi n_{g,p})\,\sigma^{z}_{p}\,/2
+
\delta h^{y}\sin(\pi n_{g,p})\,\sigma^{y}_{p}.
\label{singlequbitham}
\end{align}
From this result, we see that the $\cos(\phi_{p})$ and $\sin(\phi_{p})$ elements induce rotations around the $x$- and $y$-axis of the Bloch sphere.
The respective matrix elements are given by $\delta h^{x}=\left\langle 0_{p}| H^{x}_{}|1_{p}\right\rangle$ and $\delta h^{y}\sin(\pi n_{g,p})=\left\langle 0_{p}|H^{y}_{}|1_{p}\right\rangle$. The dependence of these matrix elements on the offset charge $n_{g,p}$ is shown in Fig.\,\ref{fig:3}(b). Since we can reach any point on the Bloch sphere through a combined rotation around the $x$- and $y$-axis, we conclude that the free time evolution of the Hamiltonians in Eq.\,\eqref{singlequbitham} can implement a complete set of single-qubit gates. However, we also emphasize that these single-qubit gates break the Cooper-pair parity conservation so that the PPQ is prone to relaxation errors during the operation time of the single-qubit gates.

\subsection{$\textsf{CNOT}$ and $\textsf{SWAP}$ gate}
We now combine the proposed method for single-qubit gates with the $\textsf{CZ}^{10}_\phi$ (with $\phi = \pi$) gate to realize a $\textsf{CNOT}_{tp}$ gate with the transmon as control and the PPQ as target by the gate sequence,
\begin{equation}
\textsf{CNOT}_{tp} = 
\begin{quantikz}[column sep=0.2cm, row sep = 0.75cm]
 & \ctrl{1} & \qw \\
 & \targ{} & \qw
\end{quantikz}
=\begin{quantikz}[column sep=0.2cm, row sep = 0.35cm]
\qw & \qw &\gate[wires=2]{\textsf{CZ}^{10}} & \qw & \qw\\
\qw &  \gate{\textsf{Y}_{\frac{\pi}{2}}}  &  & \gate{\textsf{Y}_{-\frac{\pi}{2}}} & \qw
\end{quantikz}
\end{equation}
A $\textsf{CNOT}_{pt}$ gate that uses the PPQ as control and the transmon as target is similarly given by $\textsf{CNOT}_{pt}=\textsf{H}_{t}\cdot \textsf{H}_{p} \cdot \textsf{CNOT}_{tp}\cdot \textsf{H}_{t}\cdot \textsf{H}_{p}$ with the Hadamards, $\textsf{H}_{t/p}=(\sigma^{x}_{t/p}+\sigma^{z}_{t/p})/\sqrt{2}$. Most notably, the $\textsf{CNOT}_{tp}$ and $\textsf{CNOT}_{pt}$ gate can now be combined to realize a $\textsf{SWAP}=\textsf{CNOT}_{tp}\cdot \textsf{CNOT}_{pt}\cdot \textsf{CNOT}_{tp}$ operation. The \textsf{SWAP} operation enables the \textit{coherent} transfer of quantum information between the transmon and the PPQ. Interestingly, this coherent state transfer also gives a novel read-out method for the PPQ by swapping the quantum information onto the transmon and performing the read-out on the latter.

\section{Errors on the parity-protected qubit} 
In the previous sections, we have focused on deriving a scheme for a $\textsf{CZ}_\phi$ gate within our hybrid qubit setup. For our scheme, we have assumed that the Cooper-pair parity on the PPQ is conserved during the gate operation time. An interesting question is if the gate protocol modifies if errors due to \textit{unintentional} single Cooper-pair tunneling terms, as given by Eq.\,\eqref{single-qubit-perturbations}, are present on the PPQ?

\subsection{$\boldsymbol{\sin(\phi_{p})}$ errors}
To address this question, we consider the PPQ at its $n_{g,p}=0$ operation point for optimal gate-speed. We initially consider an error term, $H^{y}_{}
= 
-\varepsilon^{y}\,\sin(\phi_{p})$, with an amplitude $\varepsilon^{y}$ that is small compared to the remaining energy scales of the setup. This $\sin(\phi_p)$ error arises in a PPQ realized by a nanowire Josephson interferometer if the two interferometers junctions are not in balance \cite{schrade2021}. 
Due to the error term, we find that the low-energy Hamiltonian of Eq.\,\eqref{lowenergy1} changes to, 
\begin{equation}
H^{(n_{g,p}=0)}_{\text{low}}
\rightarrow
\begin{pmatrix}
\omega_{11} & 0 & 0 & 0 &\lambda' & 0 \\
0 & \omega_{10}  & 0 & 0 & 0 & -\lambda'' \\
0 & 0 & \omega_{01} & 0 & 0 & \kappa \\
0 & 0 & 0 & \omega_{00} &  \kappa & 0  \\
\lambda'  & 0 & 0 & \kappa & \omega_{02} & 0 \\
0 & -\lambda''  & \kappa & 0 & 0 & \omega_{03}
\end{pmatrix}
.
\end{equation}
Here, we introduced the real-valued matrix element $\kappa=\left\langle 0_{t},1_{p}|H^{y}|0_{t},3_{p}\right\rangle=\left\langle 0_{t},0_{p}|H^{y}|0_{t},2_{p}\right\rangle$. Moreover, in accordance with Eq.\,\eqref{singlequbitham}, couplings of states with opposite Cooper-pair parity within the qubit subspace $\mathcal{P}_0$ are found to be absent at $n_{g,p}=0$, .

\begin{figure}[ht] \centering
\includegraphics[width=\linewidth]{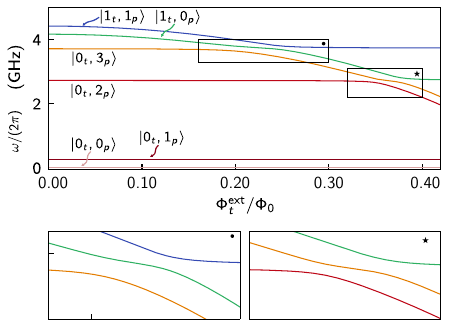}
\caption{\textbf{Effect of a cosine error in the PPQ.}
Low-energy spectrum of the hybrid qubit setup for $n_{g, p}=0$ and $(E_{J,t}, E_{J,p},E_{C,t},E_{C,p},E_{C,c}) 
= 2\pi (12,2.7,0.2,0.18,0.025)\,\text{GHz}$ as a function of the external flux $\Phi_{\text{ext}}$ of the tunable transmon in the presence of an error term $H^{x} = \varepsilon^x\,\cos(\phi_{p})$ with $\varepsilon^x=0.05 E_{J, p}$. The error term introduce additional anticrossings between states belonging to the same pair.
}\label{fig:4}
\end{figure}

Next, we integrate out the non-computational states, $\left|0_{t},2_{p}\right\rangle$ and $\left|0_{t},3_{p}\right\rangle$, with a Schrieffer-Wolff transformation
and move to the rotating frame of the bare qubit frequencies. The effective rotating frame Hamiltonian of Eq.\,\eqref{rotframeHamiltonian} then modifies to,
\begin{subequations}
\begin{align}
&\tilde{H}^{(n_{g,p}=0)}_{\text{eff}}\rightarrow
\frac{g^{zz}_{+}}{2}
\frac{
\sigma^{z}_{t}
}
{
2
}
+
\frac{g^{zz}_{-}}{2}
\frac{
\sigma^{z}_{p}
}
{
2
}
+
g^{zz}_{-}\,\frac{
\sigma^{z}_{t}
}
{
2
}
\frac{
\sigma^{z}_{p}
}
{
2
}
\\
&+(e^{i(\omega_{p}+\omega_{t})t}
g^{++}\,
\sigma^{+}_{t}\sigma^{+}_{p}
+
e^{i(\omega_{t}-\omega_{p})t}
g^{+-}\,
\sigma^{+}_{t}\sigma^{-}_{p}+\text{H.c.}),\nonumber
\end{align}
with the coefficients,
\begin{align}
g^{++}&=\frac{\kappa\lambda'}{2(\omega_{11}-\omega_{02})}, \quad g^{+-}=\frac{\kappa\lambda''}{2(\omega_{03}-\omega_{10})}.
\end{align}
\end{subequations}

It is now instructive to compare this result to the case of two capactively coupled transmons, $t1$ and $t2$, near the operation point of the $\textsf{iSWAP}$ gate \cite{krantz2019}.
In the latter case, the effective Hamiltonian comprises similar terms, $\propto \sigma^{+}_{t1}\sigma^{-}_{t2}$ and $\propto \sigma^{+}_{t1}\sigma^{+}_{t2}$, that are 
`rotating' with a factor $e^{i(\omega_{t1}-\omega_{t2})t}$ and `counter-rotating' with a factor $e^{i(\omega_{t1}+\omega_{t2})t}$, respectively. For $\omega_{t1}\approx\omega_{t2}$, the `counter-rotating' terms, which are fast-oscillating, average to zero within a `rotating-wave approximation'. Only the `rotating'  terms, which oscillate slowly, are thus retained in the effective qubit Hamiltonian. In our case, the situation is very different. Because $\omega_{t}\gg\omega_{p}$, both factors, $e^{i(\omega_{t}+\omega_{p})t}$ and $e^{i(\omega_{t}-\omega_{p})t}$, are fast-oscillating. Within a `rotating-wave approximation', we thus expect that both error terms average to zero.

To formalize this `rotating-wave approximation' argument, we integrate out the fast-oscillating terms with a time-dependent Schrieffer-Wolff transformation. To second order in $g^{++}$ and $g^{+-}$, we find that \cite{bib:supplemental},
\begin{equation}
\begin{split}
\tilde{H}^{(n_{g,p})}_{\text{eff}}
&\approx
\left(
\frac{g^{zz}_{+}}{2}
+
\frac{2[\tilde{g}^{xx}(t)-\tilde{g}^{yy}(t)]^{2}}{\omega_{t}}
\right)
\frac{
\sigma^{z}_{t}
}
{
2
}
\\
&+
\left(
\frac{g^{zz}_{-}}{2}
+
\frac{2[\tilde{g}^{xx}(t)-\tilde{g}^{yy}(t)]^{2}}{\omega_{t}}
\right)
\frac{
\sigma^{z}_{p}
}
{
2
}
\\
&+
\left(
g^{zz}_{-}
-
\frac{4[\tilde{g}^{xx}(t)+\tilde{g}^{yy}(t)]^{2}}{\omega_{t}}
\right)
\frac{
\sigma^{z}_{t}
}
{
2
}
\frac{
\sigma^{z}_{p}
}
{
2
},
\label{rotframeHamiltonian3}
\end{split}
\end{equation}
with $\tilde{g}(t)=g\sin(\omega_{t}t/2)$. From this expression for the effective rotating-frame Hamiltonian, we conclude that the mitigation of the effects of $\sin(\phi_{p})$ errors requires us to operate the setup in the regime when $g^{xx}\ll \omega_{t}$ and $g^{yy}\ll \omega_{t}$.

\subsection{$\boldsymbol{\cos(\phi_{p})}$ errors}

\begin{figure}[!t] \centering
\includegraphics[width=1.\linewidth] {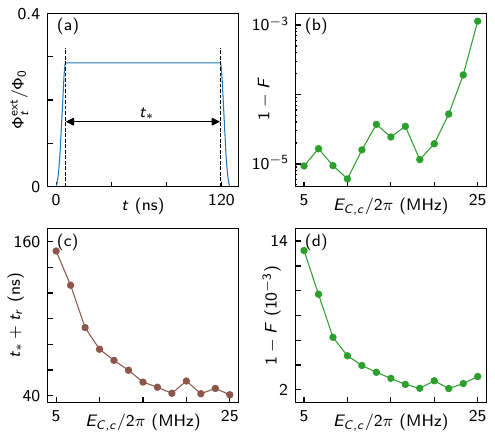}
\caption{\textbf{Performance of the $\mathbf{\textsf{CZ}^{10}_\phi}$ gate.}
(a) A typical flux pulse, $\Phi^{\text{ext}}_t(t)$, for the $\textsf{CZ}^{10}_\phi$ gate as described by Eq.\,\eqref{pulseshape}. The wait time 
near the anti-crossing $\left|1_{t},0_{p}\right\rangle\leftrightarrow\left|0_{t},3_{p}\right\rangle$ is $t_{*}$. The ramp up/down time of the flux pulse is $t_{r}/2$.
(b) Optimized gate error, $1-F$, in the absence of $1/f$ flux noise and qubit relaxation errors obtained from Eq.\,\eqref{unitaryfidelity} versus the coupling capacitance, $E_{C,c}$. 
The system parameters are $(E_{J,t}, E_{J,p},E_{C,t},E_{C,p})  
= 2\pi (12,2.7,0.2,0.15)\,\text{GHz}$  and $n_{g,p}=0$. (c) Optimized gate time, $t_{*}+t_{r}$, versus the coupling capacitance, $E_{C,c}$. The system parameters are the same as in (b). (d) Optimized gate error, $1-F$, in the presence of $1/f$ flux noise and qubit relaxation errors obtained from Eq.\,\eqref{nonunitaryfidelity}, as a function of the coupling capacitance, $E_{C,c}$. The system parameters are the same as in (b). The noise parameters are $\Gamma^{(\text{even})}_{1}=\Gamma^{(\text{odd})}_{1}=1/T_{1}$ with $T_{1}=20\,\mu$s \cite{Hutchings2017}. Moreover, $A_{1/f,\Phi}=5\,\mu\Phi_{0}$ \cite{braum2020} and $\lambda_{1/f}=4$\, \cite{KouFluxonium}.
}
\label{figure5}
\end{figure}

It is now interesting to compare our results for $\sin(\phi_p)$ errors with $\cos(\phi_p)$ errors that are described by an error term $H^{x}_{}
= 
-\varepsilon^x\,\cos(\phi_{p})$ in the Hamiltonian. Such an error term can arise in an implementation of the PPQ with a nanowire Josephson interferometer if
the external flux that threading the interferometer loop is detuned from half flux quantum \cite{schrade2021}. In this situation, the low-energy Hamiltonian of Eq.\,\eqref{lowenergy1} modifies to,
\begin{equation}
H^{(n_{g,p}=0)}_{\text{low}}
\rightarrow
\begin{pmatrix}
\omega_{11} & \delta h^{x} & 0 & 0 &\lambda' & 0 \\
\delta h^{x} & \omega_{10}  & 0 & 0 & 0 & -\lambda'' \\
0 & 0 & \omega_{01} &  \delta h^{x}  & 0 & 0 \\
0 & 0 &  \delta h^{x}  & \omega_{00} & 0 & 0  \\
\lambda'  & 0 & 0 & 0 & \omega_{02} & \chi \\
0 & -\lambda''  &0 & 0 & \chi & \omega_{03}
\end{pmatrix}
,
\end{equation}
with the matrix element $\chi=\left\langle 0_{t},2_{p}|H^{y}|0_{t},3_{p}\right\rangle$. Importantly, we see that the $\cos(\phi_p)$ errors do not lead to off-diagonal terms
that couple the matrix blocks representing the qubit subspace $\mathcal{P}_0$ and the non-computational subspace $\{|0_{t},2_{p}\rangle,|0_{t},3_{p}\rangle\}$. Consequently, we note that the $\cos(\phi_p)$ errors primarily induces mixing of opposite-parity states on the PPQ as described by $H^{x}_{\text{eff}}$ in Eq.\,\eqref{singlequbitham}. 

In summary, we have found that the nature of $\sin(\phi_p)$ errors and $\cos(\phi_p)$ errors is different in our hybrid qubit. While the $\sin(\phi_p)$ errors lead primarily to 
additional two-qubit interactions that become less relevant in the limit when $g^{xx}\ll \omega_{t}$ and $g^{yy}\ll \omega_{t}$, the $\cos(\phi_p)$ errors lead primarily to additional single-qubit terms. Finding strategies of mitigating such flux errors, for example by concatenating multiple imperfect PPQs \cite{bell2014,schrade2021}, is an important open challenge of the field.

\section{Errors on the transmon qubit} 
Besides the possible errors on the PPQ, it is essential to note that the performance of the $\textsf{CZ}^{10}_\phi$ gate in our hybrid setup can also be affected by errors on the transmon qubit. One source of such errors is $1/f$ flux noise \cite{Hutchings2017,Martinis2014}, which can give rise to fluctuations in the transmon qubit frequency, $\omega_{t}(\Phi^{\text{ext}}_t)$, and thus induce qubit dephasing. In its idle configuration at $\Phi^{\text{ext}}_t=0$, the flux-tunable transmon is always first-order protected against flux noise, $\partial\omega_{t}/\partial \Phi^{\text{ext}}_{t}=0$ at $\Phi^{\text{ext}}_t=0$. However, when tuning transmon away from $\Phi^{\text{ext}}_t$ to realize the $\textsf{CZ}^{10}_\phi$ gate it becomes susceptible to flux noise, because (in general) $\partial\omega_{t}/\partial \Phi^{\text{ext}}_{t}\neq 0$ when $\Phi^{\text{ext}}_t\neq0$. In this section, we would like to understand how our proposed $\textsf{CZ}^{10}_\phi$ gate performs in the presence of realistic $1/f$ flux noise amplitudes, which are of the order of a few $\mu\Phi_{0}$ at $1$Hz \cite{braum2020}.

As a starting point of our analysis, we assume that the PPQ is in its protected regime, as described by $H_{p}$ in Eq.\,\eqref{Eq1}, and tuned to $n_{g,p}=0$
for optimal gate speed. Following our previously outlined protocol, the $\textsf{CZ}^{10}_\phi$ gate is then realized via a rapid excursion from the idle configuration at $\Phi^{\text{ext}}_t= 0$ to a flux $\Phi^{\text{ext}}_t= \Phi^{\text{ext}}_{*}$ close to the anti-crossing $\left|1_{t},0_{p}\right\rangle\leftrightarrow\left|0_{t},3_{p}\right\rangle$ and back. We parametrize this excursion within the time interval $[0,t_{*}+t_{r}]$ through the following pulse shape \cite{YChen2021},
\begin{equation}
\Phi^{\text{ext}}_t(t)=\Phi^{\text{ext}}_{*}
\begin{cases}
 C(e^{-f(t)} -1),  &t >\frac{t_{r}}{2}+t_{*}\\
 C(e^{-f(t-t_{*})} -1),  &t <\frac{t_{r}}{2}\\
 1, & \text{else}.\\
\end{cases}
\label{pulseshape}
\end{equation}
Here, $t_{r}/2$ denotes the rise/decay time of the pulse and $t_{*}$ is the wait time near the anti-crossing. Moreover, we have introduced the constant  $C=1/(e^{-A/4}-1)$ and the function $f(t)=At(t-t_{r})/t^{2}_{r}$ with the parameter $A$ that sets the curvature of the rising/decaying pulse. An example of the pulse shape is shown in Fig.\,\ref{figure5}(a).

With the help of Eq.\,\eqref{pulseshape}, it is instructive to first look into errors of the unitary time evolution. To assess the importance of such unitary errors, we numerically solve $i\partial_{t}U(t)=\tilde{H}^{(n_{g,p}=0)}_{\text{low}}(\Phi^{\text{ext}}_t(t))U(t)$, where $\tilde{H}^{(n_{g,p}=0)}_{\text{low}}$ is represented in the rotating frame of the bare frequencies at $\Phi^{\text{ext}}_t=0$, and project the resulting time-evolution operator, $U(t_{r}+t_{*})$, onto the computational subspace. The projected operator is (in general) non-unitary due to leakage to the non-computational states and of the form $U_{c}=\text{diag}[a_{11}e^{i\phi_{11}},a_{10}e^{i\phi_{10}},1,1]$. For comparing $U_{c}$ to the $\textsf{CZ}^{10}_\phi$ gate, we apply a single-qubit $\textsf{Z}$ operation, yielding $U'_{c}=U_{\textsf{Z}}U_{c}$ with $U_{\textsf{Z}}=\text{diag}[e^{-i\phi_{11}},e^{-i\phi_{11}},1,1]$. We then use $U'_{c}$ to compute the gate fidelity \cite{Pedersen2007,xiong2022},
\begin{equation}
F=[\text{Tr}(U'^{\dagger}_{c}U'_{c})+
|
\text{Tr}(
U^{\dag}_{\textsf{CZ}^{10}_\phi}U'_{c}
)
|^{2}
]/20,
\label{unitaryfidelity}
\end{equation}
where $U_{\textsf{CZ}^{10}_\phi}=\text{diag}[1,e^{-i\phi},1,1]$. In our simulations, we have focused on $\phi=\pi$ and optimized over the pulse parameters $(\Phi^{\text{ext}}_{*},A,t_{r},t_{*})$. The resulting gate errors, $1-F$, and gate times are shown in Fig.\,\ref{figure5}(b) and (c) versus the coupling charging energy, $E_{C,c}$. While the gate time is reduced for stronger couplings, we find that the gate errors increases upon increasing $E_{C,c}$. We attribute this increase in $1-F$ to an increase in both the phase error, $e^{i(\phi_{10}-\phi_{11})}\neq -1$, and the error due to leakage to non-computational states, $a_{11}\neq1$ or $a_{10}\neq1$.

Having discussed the effect of errors in the unitary time evolution, we now proceed by analyzing the performance of the $\textsf{CZ}^{10}_\phi$ gate in the presence of incoherent errors, including $1/f$ flux noise. To determine the time-evolution of the density matrix in the presence of $1/f$ flux noise, we follow the approach of \cite{didier2019} and consider a phenomenological master equation of the form, 
\begin{equation}
\begin{split}
\partial_{t}\rho
&=[\tilde{H}^{(n_{g,p}=0)}_{\text{low}},\rho]
+\mathcal{D}[L_{1,t}]\rho+\mathcal{D}[L_{1,p}]\rho \\
&+\mathcal{D}[L^{(10)}_{\varphi}(t)]\rho+\mathcal{D}[L^{(11)}_{\varphi}(t)]\rho,
\end{split}
\label{mastereq}
\end{equation}
where $\mathcal{D}[L]\rho=L\rho L^{\dag}-(L^{\dag}L\rho+\rho L^{\dag}L)/2$.

For the time-independent collapse operators describing relaxation errors, we use,
\begin{equation}
\begin{split}
L^{(\text{even})}_{1}&=
\sqrt{\Gamma^{(\text{even})}_{1}}
(
|0_{t},0_{p}\rangle\hspace{-2pt}
\langle 1_{t}, 0_{p}|
+
|0_{t},0_{p}\rangle\hspace{-2pt}
\langle 0_{t}, 3_{p}|
\\
&\hspace{45pt}+
|0_{t},3_{p}\rangle\hspace{-2pt}
\langle 1_{t}, 0_{p}|
)
,
\\
L^{(\text{odd})}_{1}&=
\sqrt{\Gamma^{(\text{odd})}_{1}}
(
|0_{t},1_{p}\rangle\hspace{-2pt}
\langle 1_{t}, 1_{p}|
+
|0_{t},1_{p}\rangle\hspace{-2pt}
\langle 0_{t}, 2_{p}|
\\
&\hspace{42pt}+
|0_{t},2_{p}\rangle\hspace{-2pt}
\langle 1_{t}, 1_{p}|).
\end{split}
\label{CollapseOp}
\end{equation}
Here, $\Gamma^{(\text{even/odd})}_{1}$ are the decay rates within the even/odd Cooper-pair parity sectors. We have assumed, for simplicity, that within a particular Cooper-pair parity sector, all decay channels are characterized by the same decay rate. Moreover, we have assumed that due to the conservation of Cooper-pair parity on the PPQ, any decay channels that connect the two Cooper-pair parity sectors are suppressed.

For the time-dependent collapse operators accounting for $1/f$ flux noise, we use, 
\begin{equation}
\begin{split}
L^{(10)}_{\varphi}(t)&=2\sqrt{t}\,\Gamma^{(10)}_{\varphi}(t)\,|1_{t},0_{p}\rangle\hspace{-2pt}
\langle 1_{t}, 0_{p}|,
\\
L^{(11)}_{\varphi}(t)&=
2\sqrt{t}\,\Gamma^{(11)}_{\varphi}(t)\,
|1_{t},1_{p}\rangle\hspace{-2pt}
\langle 1_{t}, 1_{p}|,
\end{split}
\label{CollapseOp}
\end{equation}
with the $1/f$ flux noise dephasing rates, 
\begin{equation}
\Gamma^{(ss')}_{\varphi}(t)
= 
\lambda_{1/f} \left|
\frac{
\partial\omega_{ss'}
}{
\partial\Phi^{\text{ext}}_t
}(t)
\right|
A_{1/f,\Phi}.
\end{equation}
Here, $\lambda_{1/f}$ is a dimensionless numerical prefactor and $A_{1/f,\Phi}$ denotes the amplitude of the $1/f$ flux noise power spectral density, $S(\omega)=2\pi A^{2}_{1/f,\Phi}/|\omega|$. We have assumed, again for simplicity, that the dephasing arises primarily from the flux-dependence of the 
$|1_{t},0_{p}\rangle$ and $|1_{t},1_{p}\rangle$ levels.

To estimate the gate error, $1-F$, in the presence of $1/f$ flux noise and the decay channels, we follow closely the procedure in \cite{xiong2022}: For a given initial state, $|\psi_{0}\rangle$, we first compute the
time evolution of the density matrix, $\rho$, from Eq.\,\eqref{mastereq} using the QuTip package \cite{Johansson2012}. Subsequently, we compute the state-dependent 
gate fidelity,
\begin{equation}
F_{\rho}=\text{Tr}[\rho\rho_{\text{ideal}}].
\label{nonunitaryfidelity}
\end{equation}
Here, $\rho_{\text{ideal}}=|\psi_{\text{ideal}}\rangle\hspace{-2pt}\langle\psi_{\text{ideal}}|$ with $|\psi_{\text{ideal}}\rangle=U^{\dag}_{\textsf{Z}}U_{\textsf{CZ}^{10}_\phi}|\psi_{0}\rangle$ and $U_{\textsf{Z}}$ is obtained from the calculation of the unitary error. We repeat this procedure for 36 initial two-qubit states obtained by combining the single-qubit states $\{|0_{t/p}\rangle,|1_{t/p}\rangle,(|0_{t/p}\rangle\pm|1_{t/p}\rangle)/\sqrt{2},(|0_{t/p}\rangle\pm i|1_{t/p}\rangle)/\sqrt{2}\}$. By averaging the resulting values for $1-F_{\rho}$, we arrive at an estimate for $1-F$.

Our results from the aforementioned procedure are shown in Fig.\,\ref{figure5}(d) for a typical set of system and noise parameters. We find that the gate error depends strongly on the magnitude of the coupling charging energy, $E_{C,c}$. For smaller capacitive couplings, $E_{C,c}/2\pi = 5\,\text{MHz}$, corresponding to longer gate times, 152ns, we find a rather substantial reduction of the fidelity to $F\approx98.6$\%. In contrast, for stronger couplings, $E_{C,c}/2\pi = 20\,\text{MHz}$, the shorter gate times, 51ns, reduce the exposure to low-frequency flux noise and the decay channels. As a result, 
the theoretical gate fidelity can reach $F\approx99.7$\%, which is comparable to entangling gates between transmons \cite{kjaergaard2020}. However, we acknowledge that additional factors may further degrade the theoretical gate fidelity values in experiments. For example, it is to be expected that the effects of $1/f$ flux noise become more acute when many qubits are operated on the same chip. In this scenario, realizing accurate qubit calibration and high-fidelity gate operations will become more difficult. An interesting challenge for future works will be to further optimize gate protocols for hybrid PPQ-transmon devices, for example, by using dynamical decoupling techniques \cite{Pokharel2018} or optimal control \cite{Propson2022}.

\section{Conclusion} 
To conclude, we have proposed a coupling scheme for entangling a parity-protected superconducting qubit
with a conventional transmon qubit and discussed coherent state transfer as an application. While our scheme could open the way for using PPQs as quantum memories in a transmon architecture, it could also allow for a comparison of coherence times of the two qubits types within the same device.

\section*{Acknowledgements} 
We gratefully acknowledge stimulating discussions
with K. Flensberg and A. Gyenis. 
This work was supported by the Danish National Research Foundation, the Danish Council for Independent Research  \textbar\, Natural Sciences.
MK gratefully acknowledges support from the Villum Foundation (grant
37467) through a Villum Young Investigator grant.  We acknowledge support from the Microsoft Corporation.

\clearpage
\appendix
\begin{widetext}

\begin{center}
\large{\bf Supplemental Material to `Entangling transmons with low-frequency protected superconducting qubits' \\}
\end{center}
\begin{center}
Andrea Maiani, Morten Kjaergaard, and Constantin Schrade
\\
{\it Center for Quantum Devices, Niels Bohr Institute, University of Copenhagen, 2100 Copenhagen, Denmark}
\end{center}
In the Supplemental Material, we provide details on the derivations of the effective Hamiltonians presented in the main text. We also provide more details on the energy level structure of the hybrid qubit setup.
\section{Time-independent effective Hamiltonians}
In this section of the Supplemental Material, we give details on the derivation of the time-independent effective Hamiltonians for the coupled qubit setup as presented in 
Eq.\,(7) of the main text.
\\

As a starting point, we project the setup Hamiltonian $H$ onto $\{\left|0_{t},0_{p}\right\rangle,\left|1_{t},0_{p}\right\rangle,\left|0_{t},1_{p}\right\rangle,\left|1_{t},1_{p}\right\rangle,\left|0_{t},2_{p}\right\rangle,\left|0_{t},3_{p}\right\rangle \}$, which correspond to the relevant low-energy states of the uncoupled Hamiltonian $H_{0}$. The resulting projected Hamiltonian is given by $H^{(n_{g,p})}_{\text{low}}=H^{(0)}_{\text{low}}+H^{(1)}_{\text{low}}+H^{(2)}_{\text{low}}$ with,
\begin{equation}
H^{(0)}_{\text{low}}=
\begin{pmatrix}
\omega_{11} & 0 & 0 & 0 & 0 & 0 \\
0 & \omega_{10} & 0 & 0 & 0 & 0 \\
0 & 0 & \omega_{01}& 0 & 0& 0 \\
0 & 0 & 0 & \omega_{00} & 0 & 0 \\
0 & 0 & 0 & 0 & \omega_{02} & 0 \\
0 & 0 & 0 & 0 & 0 & \omega_{03} 
\end{pmatrix}
,
\quad
H^{(1)}_{\text{low}}=
\begin{pmatrix}
0 & 0 & -i\eta' & 0 & 0 & 0 \\
0 & 0 & 0 & i\eta'' & 0 & 0 \\
i\eta' & 0 &0& 0 & 0& 0 \\
0 & -i\eta'' & 0 & 0& 0 & 0 \\
0 & 0 & 0 & 0 &0 & 0 \\
0 & 0 & 0 & 0 & 0 & 0 
\end{pmatrix}
,
\quad
H^{(2)}_{\text{low}}=
\begin{pmatrix}
0 & 0 & 0 & 0 & \lambda' & 0 \\
0 & 0 & 0 & 0 & 0 & -\lambda'' \\
0 & 0 & 0& 0 & 0& 0 \\
0 & 0 & 0 & 0 & 0 & 0 \\
\lambda' & 0 & 0 &0 &0 & 0 \\
0 & -\lambda'' &0 & 0 & 0 & 0
\end{pmatrix}.
\end{equation}
Here we have chosen a gauge of the wavefunctions in the uncoupled system such that $\langle 2_{p}|n_{p}|1_{p}\rangle$, $\langle 3_{p}|n_{p}|0_{p}\rangle$, and $\langle 0_{t}|n_{t}|1_{t}\rangle$ are purely imaginary-valued. As a result of this gauge choice, the following quantities are real-valued,
\begin{equation}
\begin{split}
\lambda'&=\left\langle 1_{t},1_{p}|H_{c}|0_{t},2_{p}\right\rangle,
\\
-\lambda''&=\left\langle 1_{t},0_{p}|H_{c}|0_{t},3_{p}\right\rangle,
\\
 -i\eta'&=\left\langle 1_{t},1_{p}|H_{c}|0_{t},1_{p}\right\rangle,
 \\
  i\eta''&=\left\langle 1_{t},0_{p}|H_{c}|0_{t},0_{p}\right\rangle.
\end{split}
\end{equation}
\\

Next, we integrate out the effects of the non-computational states $\{\left|0_{t},2_{p}\right\rangle,\left|0_{t},3_{p}\right\rangle \}$ by means of a Schrieffer-Wolff transformation \cite{Bravyi2011_SM,Rancic2019_SM}. As the generator of the Schrieffer-Wolff transformation, we use 
\begin{equation}
S=
\begin{pmatrix}
0 & 0 & 0 & 0 & -\Omega'/\lambda' & 0 \\
0 & 0 & 0 & 0 & 0 & \Omega''/\lambda'' \\
0 & 0 & 0& 0 & 0& 0 \\
0 & 0 & 0 & 0 & 0 & 0 \\
\hline
\Omega'/\lambda'  & 0 & 0 & 0 &0 & 0 \\
0 & - \Omega''/\lambda''& 0 & 0 & 0 & 0
\end{pmatrix}
\quad
\text{with}
\quad
\Omega'=\frac{\lambda'^{2}}{\omega_{11}-\omega_{02}}\quad\text{and} \quad 
\Omega''=\frac{\lambda''^{2}}{\omega_{10}-\omega_{03}}.
\end{equation}
We note that the generator satisfies $[H^{(0)}_{\text{low}},S]=-H^{(2)}_{\text{low}}$. To second order in $\lambda'$ and $\lambda''$, the Schrieffer-Wolff generator produces an effective Hamiltonian $H^{(n_{g,p})}_{\text{eff}}=H^{(0)}_{\text{low}}+H^{(1)}_{\text{low}}+[H^{(2)}_{\text{low}},S]/2$. Evaluating this expression yields,
\begin{equation}
\begin{split}
H^{(n_{g,p})}_{\text{eff}}=
\left(\omega_{t}+\frac{g^{zz}_{+}}{2}\right)
\frac{
\sigma^{z}_{t}
}
{
2
}
+
\left(\omega_{p}+\frac{g^{zz}_{-}}{2}\right)
\frac{
\sigma^{z}_{p}
}
{
2
}
+
g^{zz}_{-}\,\frac{
\sigma^{z}_{t}
}
{
2
}
\frac{
\sigma^{z}_{p}
}
{
2
}
+
g^{y}_{t}\,\sigma^{y}_{t}
+
g^{yz}\,\sigma^{y}_{t}\sigma^{z}_{p} 
\quad \text{with} \quad
g^{zz}_{\pm}&=\Omega'\pm\Omega'',
\\
 g^{yz}&=(\eta'+\eta'')/2,\\
g^{y}_{t}&=(\eta'-\eta'')/2.
\end{split}
\end{equation}
This corresponds to the effective Hamiltonian presented in Eq.\,(7) of the main text. 

\section{Effective Hamiltonians for the time-evolution}
In this section of the Supplemental Material, we give details on the derivation of the effective Hamiltonian in Eq.\,(10) of the main text that approximates the time-evolution of our hybrid qubit system.

\subsection{Time-dependent Schrieffer-Wolff transformation}
As a starting point, we provide a brief review of the time-dependent version of the Schrieffer-Wolff transformation.
Given some time-dependent effective Hamiltonian $H(t)$, the time-dependent Schrieffer-Wolff transformation generates an effective Hamiltonian $H_{\text{eff}}(t)$ via a unitary transformation with a time-dependent generator $S(t)$ with $S(t)=-S(t)^{\dagger}$.
By using the Baker-Campbell-Haussdorf formula, we can formulate the action of the time-dependent Schrieffer-Wolff unitary transformation on the Hamiltonian $H(t)$ as, 
\begin{equation}
\begin{split}
H_{\text{eff}}(t)-i\partial_{t}
&\equiv
e^{-S(t)}
(H(t)-i\partial_{t}
)
e^{S(t)}
\\
&=
(H(t)-i\partial_{t})
+
\left[
H(t)-i\partial_{t},S(t)
\right]
+
\frac{1}{2}
\left[
\left[
H(t)-i\partial_{t},S(t)
\right],
S(t)
\right]
+
\dots
\\
&=
H(t)-i\partial_{t}
+
\left[
H(t),S(t)
\right]
-
i\dot{S}(t)
+
\frac{1}{2}
\left[
\left[
H(t),S(t)
\right],
S(t)
\right]
-
\frac{i}{2}
\left[
\dot{S}(t),
S(t)
\right]
+
\dots
\end{split}
\end{equation}
Next, we choose the time-dependent Hamiltonian to be of the specific form,
\begin{equation}
\begin{split}
H(t)&=H^{(0)}+\xi H^{(2)}(t).
\end{split}
\end{equation}
Here, $H^{(0)}$ is a time-independent unperturbed Hamiltonian and $ H^{(2)}(t)$ is a time-dependent perturbation. The parameter $\xi$ is an aid to count the order in perturbation theory and can be set to $\xi=1$ at the end of the derivation. 
Besides specifying the form of the time-dependent Hamiltonian, we also require that the generator of the time-dependent Schrieffer-Wolff transformation satisfies the following differential equation,
\begin{equation}
\begin{split}
\xi H^{(2)}(t) + \left[ H^{(0)}, S(t) \right] - i \dot{S}(t) =0.
\end{split}
\end{equation}
Using these two conditions on the time-independent Hamiltonian and the generator $S(t)$, we find that the expression for the effective Hamiltonian $H_{\text{eff}}(t)$ can be simplified to,
\begin{equation}
\begin{split}
H_{\text{eff}}(t)-i\partial_{t}
&=
H(t)-i\partial_{t}
+
\left[
H(t),S(t)
\right]
-
i\dot{S}(t)
+
\frac{1}{2}
\left[
\left[
H(t),S(t)
\right],
S(t)
\right]
-
\frac{i}{2}
\left[
\dot{S}(t),
S(t)
\right]
+
\dots
\\
&=
H^{(0)}-i\partial_{t}
+
\frac{1}{2}
\left[
\xi H^{(2)}(t)
,
S(t)
\right]
+
\frac{1}{2}
\left[
\left[
\xi H^{(2)}(t),S(t)
\right],
S(t)
\right]
+
\dots
\end{split}
\end{equation}
We now proceed by assuming that the generator $S(t)$ can expanded in a perturbative series, 
\begin{equation}
\begin{split}
S(t)&=\xi S_{1}(t)+\xi^{2} S_{2}(t) + \cdots
\end{split}
\end{equation}
Inserting this series into the expression for the effective Hamiltonian $H_{\text{eff}}$ and only retaining terms up to order $\xi^{2}$, we find that, 
\begin{equation}
\begin{split}
H_{\text{eff}}(t)-i\partial_{t}
&=
H^{(0)}-i\partial_{t}
+
\frac{\xi^{2}}{2}
\left[
 H^{(2)}(t)
,
S_{1}(t)
\right]
+
\mathcal{O}(\xi^{3})
\end{split}
\end{equation}
Finally, we set $\xi=1$ and arrive at the following form of the effective Hamiltonian, 
\begin{equation}
\begin{split}
H_{\text{eff}}(t)-i\partial_{t}
&\equiv
H^{(0)}-i\partial_{t}
+
\frac{1}{2}
\left[
 H^{(2)}(t)
,
S_{1}(t)
\right]
\end{split}
\end{equation}

\subsection{Rotating frame for the hybrid qubit setup}
We now want to apply the time-dependent Schrieffer-Wolff transformation to our hybrid qubit setup. For that purpose, it is helpful to initially move to a rotating reference frame, which is achieved by separating the full qubit entangling Hamiltonian into, 
\begin{equation}
\begin{split}
H^{(n_{g_p})}_{\text{eff}}
=H^{(0)}+H^{(2)}
\quad
\text{with}
\quad
H^{(0)}=
\left(\omega_{t}+\frac{g^{zz}_{+}}{2}\right)
\frac{
\sigma^{z}_{t}
}
{
2
}
+
\left(\omega_{p}+\frac{g^{zz}_{-}}{2}\right)
\frac{
\sigma^{z}_{p}
}
{
2
}
+
+
g^{zz}_{-}\,\frac{
\sigma^{z}_{t}
}
{
2
}
\frac{
\sigma^{z}_{p}
}
{
2
} ,
\quad 
H^{(2)}=
g^{y}_{t}\,\sigma^{y}_{t}
+
g^{yz}_{tp}\,\sigma^{y}_{t}\sigma^{z}_{p}, 
\end{split}
\end{equation}
and applying the following time-dependent unitary transformation,
\begin{equation}
U(t)=e^{i(\omega_{t}\sigma^{z}_{t}
+
\omega_{p}\sigma^{z}_{p})t/2}.
\end{equation}
This transformation yields the qubit entangling Hamiltonian in the frame that rotates at the bare qubit frequencies with components,
\begin{equation}
\begin{split}
\tilde{H}^{(0)}&=U^{\dag}(t)H^{(0)}U(t)-iU^{\dag}(t)\dot{U}(t)=\frac{g^{zz}_{+}}{2}
\frac{
\sigma^{z}_{t}
}
{
2
}
+
\frac{g^{zz}_{-}}{2}
\frac{
\sigma^{z}_{p}
}
{
2
}
+
g^{zz}_{-}\,\frac{
\sigma^{z}_{t}
}
{
2
}
\frac{
\sigma^{z}_{p}
}
{
2
} ,\\
\tilde{H}^{(2)}(t)&=U^{\dag}(t)H^{(2)}U(t)-iU^{\dag}(t)\dot{U}(t)=
g^{y}\sin(\omega_{t}t)\,\sigma^{x}_{t}
+
g^{y}\cos(\omega_{t}t)\,\sigma^{y}_{t}
+
g^{yz}\sin(\omega_{t}t)\,\sigma^{x}_{t}\sigma^{z}_{p} 
+
g^{yz}\cos(\omega_{t}t)\,\sigma^{y}_{t}\sigma^{z}_{p} .
\end{split}
\end{equation}
The rotating frame Hamiltonian can also be explicitly written in matrix form as, 
\begin{equation}
\begin{split}
\tilde{H}^{(0)}&=
\frac{1}{4}
\begin{pmatrix}
g^{zz}_{+}+2g^{zz}_{-} & 0 &0 & 0  \\
0 & g^{zz}_{+}-2g^{zz}_{-}  & 0 &0  \\
0 & 0 & -g^{zz}_{+}& 0  \\
0 & 0 & 0 & -g^{zz}_{+}
\end{pmatrix}, \\ 
\tilde{H}^{(2)}(t)&=
\begin{pmatrix}
0 & 0 &-i (g^{y}+g^{yz}) e^{i\omega_{t}t}  & 0  \\
0 & 0 & 0 &-i (g^{y}-g^{yz}) e^{i\omega_{t}t}   \\
i (g^{y}+g^{yz})e^{-i\omega_{t}t} & 0 &0& 0  \\
0 & i (g^{y}-g^{yz}) e^{-i\omega_{t}t} & 0 &0
\end{pmatrix}.
\end{split}
\end{equation}

\subsection{Time-dependent Schrieffer-Wolff transformation for hybrid qubit setup}
We now want to perform a time-dependent Schrieffer-Wolff transformation that eliminates the fast-oscillating terms, $\propto e^{\pm i\omega_{t}t}$, in $\tilde{H}^{(2)}(t)$ 
to second order in $g^{y}$ and $g^{yz}$. We, therefore, introduce the following Schrieffer-Wolff generator,   
\begin{equation}
S_{1}(t)=
\begin{pmatrix}
0 & 0 &f_{1}(t)& 0  \\
0 & 0 & 0 &f_{2}(t)  \\
-f_{1}(t)^{*} & 0 &0& 0  \\
0 &-f_{2}(t)^{*}   & 0 &0
\end{pmatrix}
\end{equation}
with the functions,
\begin{equation}
f_{1}(t)=\frac{
-2i(g^{y}+g^{yz})
(
e^{-i(g^{zz}_{-}+g^{zz}_{+})t/2}
-
e^{i\omega_{t}t}
)
}{g^{zz}_{-}+g^{zz}_{+}+2\omega_{t}}, \quad
f_{2}(t)=\frac{
2i(g^{y}-g^{yz})
(
e^{i(g^{zz}_{-}-g^{zz}_{+})t/2}
-
e^{i\omega_{t}t}
)
}{g^{zz}_{-}-g^{zz}_{+}-2\omega_{t}}
\end{equation}
This generator satisfies,
\begin{equation}
S_{1}(t)=-S_{1}(t)^{\dagger}, \quad
 \tilde{H}^{(2)}(t) + \left[ \tilde{H}^{(0)}, S_{1}(t) \right] - i \dot{S}_{1}(t) =0,
 \quad
 \text{and} \quad
 S_{1}(0)=0.
\end{equation}
Moreover, the generator allows us to compute the effective correction term to $\tilde{H}^{(0)}(t)$ to second order in $g^{y}$ and $g^{yz}$,
\begin{equation}
\begin{split}
\frac{1}{2}
\left[
\tilde{H}^{(2)}(t),S_{1}(t)
\right]
=
\begin{pmatrix}
h_{1}(t)& 0 &0 & 0  \\
0 & h_{2}(t) & 0 &0  \\
0 & 0 & -h_{1}(t) & 0  \\
0 & 0 & 0 & -h_{2}(t) 
\end{pmatrix},
\end{split}
\end{equation}
with the functions,
\begin{equation}
\begin{split}
&h_{1}(t) = \frac{4(g^{y}+g^{yz})^{2}\sin([g^{zz}_{-}+g^{zz}_{+}+2\omega_{t}]t/4)^{2}}{g^{zz}_{-}+g^{zz}_{+}+2\omega_{t}},
\quad
h_{2}(t) = -\frac{4(g^{y}-g^{yz})^{2}\sin([g^{zz}_{-}-g^{zz}_{+}-2\omega_{t}]t/4)^{2}}{g^{zz}_{-}-g^{zz}_{+}-2\omega_{t}}
\end{split}
\end{equation}
Provided that $\omega_{t}\gg g^{zz}_{\pm}$, we neglect the terms in the denominators and sine functions that are $\propto g^{zz}_{\pm}$. When then add the correction term to 
$\tilde{H}^{(0)}(t)$, which yields the full effective Hamiltonian,
\begin{equation}
\begin{split}
\tilde{H}^{(n_{g,p})}_{\text{eff}}(t)&\approx
\left(
\frac{g^{zz}_{+}}{2}
+
\frac{4[(g^{y})^{2}+(g^{yz})^{2}]\sin(\omega_{t}t/2)^{2}}{\omega_{t}}
\right)
\frac{
\sigma^{z}_{t}
}
{
2
}
+
\frac{g^{zz}_{-}}{2}
\frac{
\sigma^{z}_{p}
}
{
2
}
+
\left(
g^{zz}_{-}
+
\frac{16g^{y}g^{yz}\sin(\omega_{t}t/2)^{2}}{\omega_{t}}
\right)
\,\frac{
\sigma^{z}_{t}
}
{
2
}
\frac{
\sigma^{z}_{p}
}
{
2
}.
\end{split}
\end{equation}
This concludes our derivation of the effective Hamiltonian for the time-evolution of our hybrid qubit setup.

\section{More details on the possible errors}
In this section of the Supplemental Material, we provide details on the derivation of the effective Hamiltonians presented in Eq.\,(15) and Eq.\,(16) of the main text. These Hamiltonians account for the presence of $\sin(\phi_p)$ error terms in our hybrid qubit setup. 
\\

First, we note that the derivations for the effective Hamiltonians \textit{with} the $\sin(\phi_p)$ error terms are very similar to the derivations for the effective Hamiltonians \textit{without} $\sin(\phi_p)$ the error terms. Since the latter derivations have been discussed in great detail in the previous sections of the Supplemental Material, we will focus only on the main modifications.

\subsection{Time-independent effective Hamiltonian}
For deriving the time-independent effective Hamiltonian of Eq.\,(15), we note that the low-energy Hamiltonian at $n_{g,p}=0$ is given by
$H^{(n_{g,p}=0)}_{\text{low}}=H^{(0)}_{\text{low}}+H^{(2)}_{\text{low}}$ with,
\begin{equation}
H^{(0)}_{\text{low}}=
\begin{pmatrix}
\omega_{11} & 0 & 0 & 0 & 0 & 0 \\
0 & \omega_{10} & 0 & 0 & 0 & 0 \\
0 & 0 & \omega_{01}& 0 & 0& 0 \\
0 & 0 & 0 & \omega_{00} & 0 & 0 \\
0 & 0 & 0 & 0 & \omega_{02} & 0 \\
0 & 0 & 0 & 0 & 0 & \omega_{03} 
\end{pmatrix}
,
\quad
H^{(2)}_{\text{low}}=
\begin{pmatrix}
0 & 0 & 0 & 0 & \lambda' & 0 \\
0 & 0 & 0 & 0 & 0 & -\lambda'' \\
0 & 0 & 0& 0 & 0& \kappa \\
0 & 0 & 0 & 0 & \kappa & 0 \\
\lambda' & 0 & 0 &\kappa &0 & 0 \\
0 & -\lambda'' &\kappa & 0 & 0 & 0
\end{pmatrix}.
\end{equation}
Here, the matrix element $\kappa=\left\langle 0_{t},1_{p}|H^{y}|0_{t},3_{p}\right\rangle=\left\langle 0_{t},0_{p}|H^{y}|0_{t},2_{p}\right\rangle$ is real-valued (in the same gauge as the one used in the first section of the Supplemental Material) and accounts for the presence of the $\sin(\phi_p)$ error terms. 
\\
\\
Next, we write down the generator of the Schrieffer-Wolff transformation,
\begin{equation}
S=
\begin{pmatrix}
0 & 0 & 0 & 0 & -\Omega'/\lambda' & 0 \\
0 & 0 & 0 & 0 & 0 & \Omega''/\lambda'' \\
0 & 0 & 0& 0 &0&  \Gamma' \\
0 & 0 & 0 & 0 & \Gamma'' & 0 \\
\Omega'/\lambda'  & 0 & 0 & -\Gamma'' &0 & 0 \\
0 & - \Omega''/\lambda''& -\Gamma' & 0 & 0 & 0
\end{pmatrix}
\quad
\text{with}
\quad
\Gamma'=\frac{\kappa}{\omega_{03}-\omega_{01}}\quad\text{and} \quad 
\Gamma''=\frac{\kappa}{\omega_{02}-\omega_{00}},
\end{equation}
The generator satisfies $[H^{(0)}_{\text{low}},S]=-H^{(2)}_{\text{low}}$ and yields the effective Hamiltonian, $H^{(n_{g,p}=0)}_{\text{eff}}=H^{(0)}_{\text{low}}+[H^{(2)}_{\text{low}},S]/2$. Projected onto the qubit subspace $\mathcal{P}_0$, the effective Hamiltonian evaluates to, 
 \begin{equation}
\begin{split}
H^{(n_{g,p}=0)}_{\text{eff}}\approx
\left(\omega_{t}+\frac{g^{zz}_{+}}{2}\right)
\frac{
\sigma^{z}_{t}
}
{
2
}
+
\left(\omega_{p}+\frac{g^{zz}_{-}}{2}\right)
\frac{
\sigma^{z}_{p}
}
{
2
}
+
g^{zz}_{-}\,\frac{
\sigma^{z}_{t}
}
{
2
}
\frac{
\sigma^{z}_{p}
}
{
2
}
+
g^{xx}_{}\,\sigma^{x}_{t}\sigma^{x}_{p} 
+
g^{yy}\,\sigma^{y}_{t}\sigma^{y}_{p} 
\end{split}
\end{equation}
with the coefficients,
\begin{equation}
g^{xx}_{}=
\frac{\kappa}{4}
\left(
\frac{\lambda'}{\omega_{11}-\omega_{02}}
+
\frac{\lambda''}{\omega_{03}-\omega_{10}}
\right)
,
\quad
 g^{yy}=\frac{\kappa}{4}
\left(
\frac{\lambda'}{\omega_{02}-\omega_{11}}
+
\frac{\lambda''}{\omega_{03}-\omega_{10}}
\right).
\end{equation}
Here, we have dropped terms $\propto 1/(\omega_{01}-\omega_{03})$  and $\propto 1/(\omega_{00}-\omega_{02})$ due to the large separation of the respective energy levels. 

\subsection{Time-dependent effective Hamiltonian}
For deriving the time-dependent effective Hamiltonian of Eq.\,(16), we transform the effective Hamiltonian, $H^{(n_{g,p}=0)}_{\text{eff}}$, to the frame that rotates with the bare qubit frequencies. The rotating-frame Hamiltonian is of the form $\tilde{H}^{(0)}+\tilde{H}^{(2)}(t)$ with the two contributions,
\begin{equation}
\begin{split}
\tilde{H}^{(0)}&=
\frac{1}{4}
\begin{pmatrix}
g^{zz}_{+}+2g^{zz}_{-} & 0 &0 & 0  \\
0 & g^{zz}_{+}-2g^{zz}_{-}  & 0 &0  \\
0 & 0 & -g^{zz}_{+}& 0  \\
0 & 0 & 0 & -g^{zz}_{+}
\end{pmatrix}, \\ 
\tilde{H}^{(2)}(t)&=
\begin{pmatrix}
0 & 0 &0  &  (g^{xx}-g^{yy}) e^{i(\omega_{t}+\omega_{p})t}  \\
0 & 0 & (g^{xx}+g^{yy}) e^{i(\omega_{t}-\omega_{p})t} &0   \\
0& (g^{xx}+g^{yy}) e^{-i(\omega_{t}-\omega_{p})t}  &0& 0  \\
 (g^{xx}-g^{yy}) e^{-i(\omega_{t}+\omega_{p})t}   &0& 0 &0
\end{pmatrix}.
\end{split}
\end{equation}
\\
\\
We now introduce the generator of the time-dependent Schrieffer-Wolff transformation,
\begin{equation}
S_{1}(t)=
\begin{pmatrix}
0 & 0 &0& f_{1}(t)  \\
0 & 0 & f_{2}(t) &0  \\
0 & -f_{2}(t)^{*} &0& 0  \\
-f_{1}(t)^{*} &0   & 0 &0
\end{pmatrix},
\end{equation}
with the functions,
\begin{equation}
f_{1}(t)=\frac{
2(g^{xx}-g^{yy})
(
e^{-i(g^{zz}_{-}+g^{zz}_{+})t/2}
-
e^{i(\omega_{t}+\omega_{p})t}
)
}{g^{zz}_{-}+g^{zz}_{+}+2(\omega_{p}+\omega_{t})}, 
\quad
f_{2}(t)=-\frac{
2(g^{xx}+g^{yy})
(
e^{i(g^{zz}_{-}-g^{zz}_{+})t/2}
-
e^{i(\omega_{t}-\omega_{p})t}
)
}{g^{zz}_{-}-g^{zz}_{+}+2(\omega_{p}-\omega_{t})}.
\end{equation}
This generator satisfies,
\begin{equation}
S_{1}(t)=-S_{1}(t)^{\dagger}, \quad
 \tilde{H}^{(2)}(t) + \left[ \tilde{H}^{(0)}, S_{1}(t) \right] - i \dot{S}_{1}(t) =0,
 \quad
 \text{and} \quad
 S_{1}(0)=0.
\end{equation}
The generator yields the effective Hamiltonian,
$\tilde{H}^{(n_{g,p}=0)}_{\text{eff}}(t)=
\tilde{H}^{(0)}+
[
\tilde{H}^{(2)}(t),S_{1}(t)
]/2$, which evaluates to,
 \begin{equation}
\begin{split}
\tilde{H}^{(n_{g,p}=0)}_{\text{eff}}(t)&\approx
\left(\frac{g^{zz}_{+}}{2}
+
\frac{2(g^{xx}-g^{yy})^{2}\sin(\omega_{t}t/2)^{2}}{\omega_{t}}
\right)
\frac{
\sigma^{z}_{t}
}
{
2
}
+
\left(\frac{g^{zz}_{-}}{2}
+
\frac{2(g^{xx}-g^{yy})^{2}\sin(\omega_{t}t/2)^{2}}{\omega_{t}}
\right)
\frac{
\sigma^{z}_{p}
}
{
2
}
\\
&+
\left(
g^{zz}_{-}
-
\frac{4(g^{xx}+g^{yy})^{2}\sin(\omega_{t}t/2)^{2}}{\omega_{t}}
\right)
\,\frac{
\sigma^{z}_{t}
}
{
2
}
\frac{
\sigma^{z}_{p}
}
{
2
}.
\end{split}
\end{equation}
This concludes our derivation of the effective Hamiltonians presented in Eq.\,(15) and Eq.\,(16) of the main text. 

\section{Numerical Schrieffer-Wolff transformation}
In this section of the Supplemental Material, we provide more details on the theory and application of the Schrieffer-Wolff transformation for the coupled qubits problem. The numerical Schrieffer-Wolff method is used in the text to derive the 6-levels effective Hamiltonians integrating out the effect of high-energy levels.

Let us consider the Hamiltonian of the coupled system $H = H_0 + H_c$ with $H_0 = H_t \otimes \mathbb{I} + \mathbb{I} \otimes H_p$ being the decoupled system Hamiltonian and $H_c$ the capacitive coupling Hamiltonian. The Hilbert space of the system can be decomposed as $\mathcal{H} = \mathcal{P}_{0} \oplus \mathcal{Q}_0 = \mathcal{P} \oplus \mathcal{Q}$ where $\mathcal{P}_0$ and $\mathcal{P}$ are the low energy Hilbert spaces of the decoupled and coupled system. The computational space of the system is identified by $\mathcal{P}_0$. For this reason, we are interested in finding a unitary $U \in \mathrm{End}(\mathcal{H})$ that maps the low-energy subspace of the interacting Hamiltonian $\mathcal{P}$ to the one of the uncoupled one $\mathcal{P}_{0}$. In other words, defining $P$ and $P_0$ the orthogonal projectors on the low-energy susbspaces
\begin{align}
    P = \sum_{i}^{d} \ket{\psi_i}\bra{\psi_i} \\
    P_0 = \sum_{i}^{d} \ket{\psi_i^0}\bra{\psi_i^0}
\end{align}
where $\ket{\psi_i}$ and $\ket{\psi_i^0}$ are, respectively, the coupled and decoupled system eigenstates and $d=4$, we are seeking a unitary $U$ that satisfies
\begin{equation}
    U P U^\dag = P_0 \qquad \Rightarrow \qquad U P = P_0 U \,.
\end{equation}
This is achieved by the Schrieffer-Wolff transformation~\cite{Bravyi2011}. 

One way to see this is as a direct rotation that can be written as the square root of the product of the two reflections
\begin{equation}
    U = \sqrt{\mathcal{M}_{\mathcal{P}_0}\mathcal{M}_{\mathcal{P}}} = \sqrt{(P_0 - Q_0)(P - Q)} = \sqrt{(2 P_0 - \mathbb{I})(2 P - \mathbb{I})}  
\end{equation}
where $Q$ and $Q_0$ are the orthogonal projectors to the high-energy subspaces and $\mathcal{M}_j$ are the reflections upon the lower energy subspaces.

The low energy Hamiltonian is then:
\begin{equation}
    H_\mathrm{eff} = P_0 U H U^\dag P_0 = U P H P U^\dag 
\end{equation}

An efficient way to tackle the problem numerically is the procedure developed in \cite{Consani2020_SM,Hita-Perez2021_SM} that we will now discuss. Since the final objective is the effect of the Schrieffer-Wolff transformation only in the low-energy sector, we can focus on the following operator product 
\begin{equation}
    P_0 U = U P = \sum_{ij} A_{ij} \ket{\psi_i^0}\bra{\psi_j} = A
\end{equation}
where $A \in \mathrm{Hom}\qty(\mathcal{P}_0, \mathcal{P})$ is a rank $d$ operator. For later use, we introduce also the rank $d$ operator $B \in \mathrm{Hom}\qty(\mathcal{P}_0, \mathcal{P})$ defined as $B = P_0 P$. Since both the operators belongs to $\mathrm{Hom}\qty(\mathcal{P}_0, \mathcal{P})$, $P_0$ acts as a left identity and $P$ is a right identity. Moreover, $A$ and $B$ are related, indeed
\begin{equation}
    (U P)^2 = (P_0 A P)^2 = P_0 A P P_0 A P = P_0 A B^\dag A P = A B^\dag A 
\end{equation}
at the same time 
\begin{equation}
    ( U P)^2 = P_0 U^2 P = P_0 (P_0 - Q_0)(P - Q) P = P_0 P = B
\end{equation}
and therefore $A B^\dag A = B$. Using singular values decomposition, we can decompose $B = W \Sigma V^\dag$ where $W$ and $V$ are unitaries and $\Sigma$ is a diagonal matrix. The equation $A B^\dag A = B$ is then solved by $A = W V^\dag$.

In a practical implementation, by starting with the Hamiltonian $H_0$ and $H$ in whatever basis, we can calculate the incomplete low-energy $d$-dimensional orthogonal eigenbasis $V_0$ and $V$ ($n \times d$ matrices) with eigenvalue matrices $W_0$ and $W$ ($d \times d$ diagonal matrices) by using the Lanczos algorithm. With these, we can calculate $B = V_0^\dag V$, that is a dimension $d$ matrix and perform SVD to calculate the unitary $A$. The effective low-energy Hamiltonian in the computational basis is then $H_\mathrm{eff} = A W A^\dag$. The drawback of this method is that we lose the information on the dressed states that is encoded in the matrix $U$.

\subsection{Smooth gauge for parametric sweeps}

The purpose of the Schrieffer-Wolff transformation is to derive an effective Hamiltonian $H_\mathrm{eff}$ written in the basis of the unperturbed system $H_0$. Since $H_\mathrm{eff}$ is not an observable of the system, the effective Hamiltonian derived is not unique but it depends on the choice of the gauge for the unperturbed eigenstates $\qty{\ket{\psi_i^0}}$. This means that it is crucial to fix a smooth gauge when sweeping over a parameter that appear also in the unperturbed Hamiltonian $H_0$. Notice that, when numerical diagonalization is employed, control over the global phase of the eigenvectors is not guaranteed. Therefore a smooth gauging algorithm needs to be applied after the diagonalization.

In the case addressed in this paper, we are free to fix the gauge of the two qubits independently since they are decoupled in $H_0$. In the case the charge offsets $n_{g,t}$ and $n_{g,p}$ are both zero, it is possible to fix the gauge consistently by imposing that the wavefunction is real at a reference point. A convenient choice is represented  by picking $\phi=0$ for $\ket{0_p}$, $\ket{1_p}$ and $\ket{0_t}$ and $\phi = \pi/4$ for $\ket{1_t}$. 

To keep a smooth gauge fixing during the sweep of the offset charge, we use a smooth gauge fixing procedure. We first discretize the $n_{g, i}$ axes in a set of $N$ points in the interval $\qty[0, k]$ where $k$ is $1$ for the transmon and $2$ for the parity qubit. We will assume an homogeneous discretization with inter-site distance $\Delta n = k / (N-1)$ for simplicity. First, an arbitrary gauge fixing, like the one described in the previous paragraph, is applied for the wavefunction at point $i=0$. Next, for each wavefunction we impose that the overlap integral with the previous point is real. In other words, we calculate the fixing phase $\beta_i$ as 
\begin{equation}
    \beta_{n, i}  = \sum_{j=0}^{i-1} \Im \ln \bra{\psi_{n, j}^0}\ket{\psi_{n, j+1}^0}
\end{equation}
and then the wavefunctions are updated as $\ket{\tilde{\psi}^0_{n,i}} = e^{-i \beta_{n,i}} \ket{\tilde{\psi}^0_{n,i}}$. This is possible because we have assumed that the index $n$ identifies corresponding eigenstates at different indexes $i$. In other words, if the index $n$ orders the states by energy we are assuming that there are no level crossings in the charge-Brillouin zone. This is not true for the parity qubit, but in this case we have labeled by $n=0,1$ the even and odd lowest states that can be identified, for example, by comparing the amplitude at $\phi=0$. 

A more general method is available for the case when it is not possible to easily identify corresponding eigenstates at different values of the parameters. In that case, the application of an additional unitary point by point is necessary.

\section{Additional results on the energy levels}
In this section of the Supplemental Material, we provide additional details on the energy spectrum of the two qubit system and design principles of an hybrid qubit. 

We recall that, for the transmon, it is possible to approximate the eigenvalues distribution by expanding the transmon Hamiltonian to fourth order. In this way, the Hamiltonian is mapped to a quantum Duffing oscillator \cite{koch2007}. This gives for the transmon the following approximate spectrum:
\begin{equation}
    E_{t, m} = - E_{J, t} + \sqrt{8 E_{C, t} E_{J, t}}\qty(m + 1/2) - \frac{E_{C, t}}{12}\qty(6 m^2 + 6 m + 3)\,.
\end{equation}

A similar approach can be used to model the parity-protected qubit in the transmon regime by treating it as a double Duffing oscillator. We can expand the potential in the two wells as 
\begin{equation}
    V(\phi) = E_{J, p} \cos(2 \phi - \pi/4)  = -E_{J, p} + 4 E_{j, p} \frac{(\phi-\phi_l)^2}{2} - 16 E_{j, p} \frac{(\phi-\phi_l)^4}{24} + o(\phi^5) 
\end{equation}
with $\phi_L = \pm \pi/4$. Next we introduce a hopping amplitude between the two wells. For semplicity, we consider allowed only hoppings between the same energy levels in the left and right well, i.e. $t_{mm'} = t_m \delta_{m m'}$. Therefore, the approximate Hamiltonian is
\begin{equation}
    H = \sum_{l=L, R} \qty[\omega_p a^\dag_l a_l - E_{j, p} - \frac{E_{C, p}}{3} (a_l + a_l^\dag)^4] + \sum_{m} \qty[ t_m \qty(1 + e^{i 2 \pi n_{g,p}})( a^\dag_L)^m(a_R)^m]
\end{equation}
where $\omega_p = 2 \sqrt{8 E_{j, p} E_{C, p}}$.
The eigenvalues of the Hamiltonian can be calculated by first order perturbation theory using the number basis $\qty{\ket{m_L m_R}}$. 

The double Duffing oscillator spectrum is composed by pairs of states located around the mean value
\begin{equation}
    \mu_{n, n+1} = \frac{E_{n} + E_{n+1}}{2} \simeq - E_{J, p} + 2 \sqrt{8 E_{C, p} E_{J, p}}\qty(\frac{n}{2} + 1/2) - 4 \frac{E_{C, p}}{12}\qty(6 \qty(\frac{n}{2})^2 + 6 \frac{n}{2} + 3)\,,
\end{equation}
for even $n$, with a splitting $\delta_{n, n+1} = E_{n+1}-E_{n} \simeq \frac{t_n}{2} \cos(\pi n_{g, p})$. Each state belong to either the even or odd parity sector. In the regime $-0.5< n_{g, p} < 0.5$ the order of the states is \textit{even}, \textit{odd}, \textit{odd}, \textit{even}, \dots, while in the regime $0.5< n_{g, p} <1.5$ is \textit{odd}, \textit{even}, \textit{even}, \textit{odd}, \dots. In the PPQ the $E_{J, p} / E_{C, p}$ ratio has a twofold role. On the one hand higher ratios reduce the splitting between pair of states (\text{i.e}, $\omega_{p, 1}$ and $\delta\omega_{p, 23} = \omega_{p, 2} - \omega_{p,3}$) on the other hand increase the separation between the pairs of states (\textit{i.e.} $\mu_{p, 23}$).

\begin{figure}[!h]
    \centering
    \includegraphics[width=\linewidth]{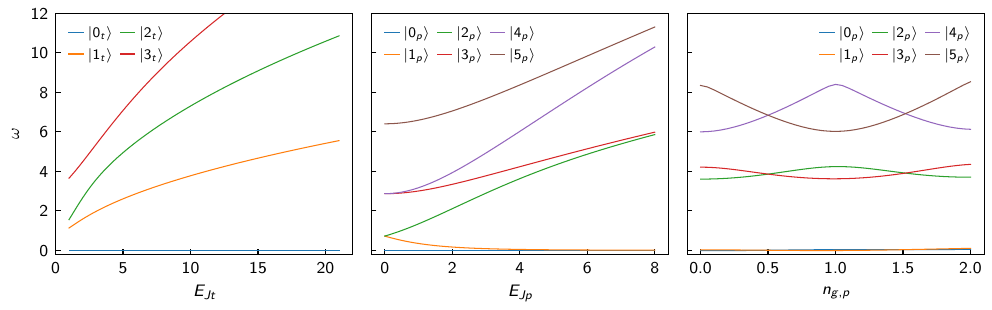}
    \caption{Low energy spectrum of: (a) a transmon with $E_{C, t} = (2\pi) \SI{0.2}{\giga\Hz}$ and (b) a parity-protected qubit with $E_{C, p} = (2\pi)\SI{0.18}{\giga\Hz}$ as function of the Josephson energy and zero offset charge. Panel (c): low energy spectrum of a parity protected qubit as a function of the offset charge with parameters $(E_{J, p}, E_{C, p}) = (2 \pi) (\SI{2.7}{\giga\Hz}, \SI{0.18}{\giga\Hz}) $}
    \label{fig:parameter_sweeps}
\end{figure}

At the optimal point $n_{g, p}=0$, states $\vert 0_t, 2_p\rangle$ belongs to the odd sector while $\vert 0_t, 3_p\rangle$ belongs to the even sector. This pair of states show a splitting $\propto \abs{\cos (\pi n_{g, p})}$ and the mean is located approximately at energy $\mu_{p, 23} = \qty(\omega_{p, 2} + \omega_{p,3})/2 \simeq 2 \sqrt{8 E_{C, p} E_{J, p}} - 4 E_{C, p}$. Depending on the parameters of the system, the pair of states can be placed above the $\vert 1_t, 0_p\rangle$ $\vert 1_t, 1_p\rangle$ pair (when $\mu_{p, 23}  \gtrsim \omega_{t, 1}$) or below (when $\mu_{p, 23} \lesssim \omega_{t, 1}$). In the presence of sizable splitting of the pair, the situation in which one of the two states lies below and one above is also possible. To obtain a controllable coupling, it is needed that at least one of the two excited states lays below the pair of computational states. For this reason, the choice of the parameters is crucial. For this reason, the approximate condition 
\begin{equation}
\sqrt{8 E_{C, t} E_{J, t}} - E_{C, t} > 2 \sqrt{8 E_{C, p} E_{J, p}} - 4 E_{C, p}
\end{equation}
has to be satisfied.

\end{widetext}
\end{document}